\documentclass[review]{elsarticle}
\usepackage{lineno,hyperref}
\modulolinenumbers[5]
\usepackage{booktabs}

\makeatletter
\def\ps@pprintTitle{%
 \let\@oddhead\@empty
 \let\@evenhead\@empty
 \def\@oddfoot{\centerline{\thepage}}%
 \let\@evenfoot\@oddfoot}
\makeatother
 
\usepackage{amssymb}
\usepackage{amsmath}
\usepackage{epsfig}
\usepackage{graphicx}
\usepackage{dcolumn}
\usepackage{array}
\usepackage{bm}
\usepackage{fancyheadings}
\usepackage{longtable}
\usepackage{multirow}
\usepackage{float}
\usepackage{afterpage}
\usepackage{subfigure}
\usepackage{mathtools}
\usepackage[monochrome]{color} 
\setlongtables

\usepackage{pbox,calc}

\usepackage{setspace}
\usepackage{etoolbox}

\usepackage[margin=2.5cm]{geometry}

\begin{document}

\begin{frontmatter}

\title{Bayesian updating for data adjustments and multi-level uncertainty propagation within Total Monte Carlo}


\author[label1]{E. Alhassan}
\ead{erwin.alhassan@psi.ch}
\author[label1]{D. Rochman}
\ead{dimitri-alexandre.rochman@psi.ch}
\author[label2]{H. Sj\"ostrand}
\author[label1]{A. Vasiliev}
\author[label2,label3]{A.J. Koning}
\author[label1]{H. Ferroukhi}

\address[label1]{Laboratory for Reactor Physics and Thermal-Hydraulics, Paul Scherrer Institut, 5232 Villigen, Switzerland}

\address[label2]{Division of Applied Nuclear Physics, Department of Physics and Astronomy, Uppsala University, Uppsala, Sweden}

 \address[label3]{Nuclear Data Section, International Atomic Energy Commission (IAEA), Vienna, Austria}

\begin{abstract}
In this work, a method is proposed for combining differential and integral benchmark experimental data within a Bayesian framework for nuclear data adjustments and multi-level uncertainty propagation using the Total Monte Carlo method. First, input parameters to basic nuclear physics models implemented within the state of the art nuclear reactions code, TALYS, were sampled from uniform distributions and randomly varied to produce a large set of random nuclear data files. Next, a probabilistic data assimilation was carried out by computing the likelihood function for each random nuclear data file based first on only differential experimental data (1st update) and then on integral benchmark data (2nd update). The individual likelihood functions from the two updates were then combined into a global likelihood function which was used for the selection of the final 'best' file. The proposed method has been applied for the adjustment of $^{208}$Pb in the fast neutron energy region below 20 MeV. The 'best' file from the adjustments was compared with available experimental data from the EXFOR database as well as evaluations from the major nuclear data libraries and found to compare favourably. 
\end{abstract}


\end{frontmatter}

\section{Introduction}
 \label{data_adjustments}
Nuclear reactors like many other technical systems are complex in nature and therefore require the coupling of several mathematical models and sub-models in order to describe and analyse them. The models used for the analysis of nuclear reactor systems can broadly be classified into the following categories: (1) Basic nuclear physics models used for the computation of basic physical observables such as nuclear reaction cross sections and angular distributions, (2) models implemented in nuclear data processing codes such as NJOY~\cite{MacFarlane-2010} and PREPRO~\cite{prepro-2012}, (3) models for neutron transport including reactor kinetics, (4) models implemented in thermal-hydraulics and computational fluid dynamic codes, (5) reactor fuel mechanics models, and (6) reactor dosimetry models,  among others. These models interact with each other since the output of some models (lower-level models) are normally used as input to other models (higher-level models). Also, feedbacks from higher-level models are normally given in order to improve lower-level models. Furthermore, different experimental data sets are utilized for the calibration and validation of these models at each level. This makes it particularly difficult and computationally expensive to integrate or combine the various activities over the entire calculation chain into a single process. With the increase in computational power and the improvements in nuclear reaction theory, a novel approach called 'Total Monte Carlo' (TMC) was developed and presented in Ref.~\cite{Koning-2008} with the goal of propagating uncertainties from basic physics parameters to applications. While this goal has been achieved to a large extent, the explicit inclusion of experimental data from both differential and integral benchmark data for combined data adjustments is still outstanding. This work seeks to combine these experimental data (differential and integral benchmark data) within a Bayesian framework, for nuclear data adjustments in the \emph{fast neutron} energy region \emph{(below 20 MeV)} using the Total Monte Carlo (TMC) method.

Within the TMC method, data adjustments to fit selected integral benchmark experimental data has been presented for example in Refs.~\cite{Rochman-2012evaluation,Roch-2011a,Rochman-2018adjustments,Rochman-2015improving,Rochman-2017correlation} and used for uncertainty reductions in reactor applications~\cite{Alhassan-2014selecting,Alhassan-2013a,Alhassan-2015NDreduction,Alhassan-2015PhD}. In Refs.~\cite{Rochman-2012evaluation,Roch-2011a}, a method that combines evaluations with a posteriori adjustments to integral measurements referred to as the 'Petten method' was developed and utilized. A random search was performed to identify the best possible nuclear data file using a goodness of fit estimator. \textcolor{blue}{The 'Petten method' was applied for the adjustment and evaluation of neutron induced reactions on $^{239}$Pu~\cite{Rchman-2011randomlyPu} and $^{63,65}$Cu~\cite{Rochman-2012evaluation} as well as for the improvement of H in $H_2O$ neutron thermal scattering data~\cite{Rochman-2012improvingH}.} In Ref.~\cite{Alhassan-2015NDreduction}, similar to Refs.~\cite{Roch-2011a,Rochman-2018adjustments,Rochman-2015improving,Rochman-2017correlation}, only information from benchmark experiments were used; differential experimental data were not explicitly incorporated into model calculations. 

Also, the inclusion of differential experimental data within the Total Monte Carlo (TMC) method has been presented previously in Refs.~\cite{Helgesson-2017uncertainty,Duan-2013,helgesson-2014incorporating,Helgesson-2017combining,koning-2015bayesianfull}. In Refs.~\cite{helgesson-2014incorporating,Helgesson-2017combining,koning-2015bayesianfull}, similar to this work, weights equal to the likelihood function were assigned to each random nuclear data file based on only differential experimental data. In Ref.~\cite{Duan-2013}, differential experimental data were incorporated into model calculations by computing a weighted $\chi^2$ for each reaction channel. In Refs.~\cite{Helgesson-2017uncertainty,Duan-2013,helgesson-2014incorporating,Helgesson-2017combining,koning-2015bayesianfull}, only information from differential experimental data were utilized in the adjustments. 

In this work, we present a method for combining experimental data from both differential and integral measurements within a Bayesian framework for data adjustments. The proposed method is applied for the adjustment of n+$^{208}$Pb cross sections in the fast energy region.

\section{Theory and Methods}

\subsection{Updating scheme}
\label{updating_scheme}
\begin{figure}[h!] 
  \centering
   \includegraphics[trim = 0mm 3mm 0mm 0mm, clip, width=0.7\textwidth]{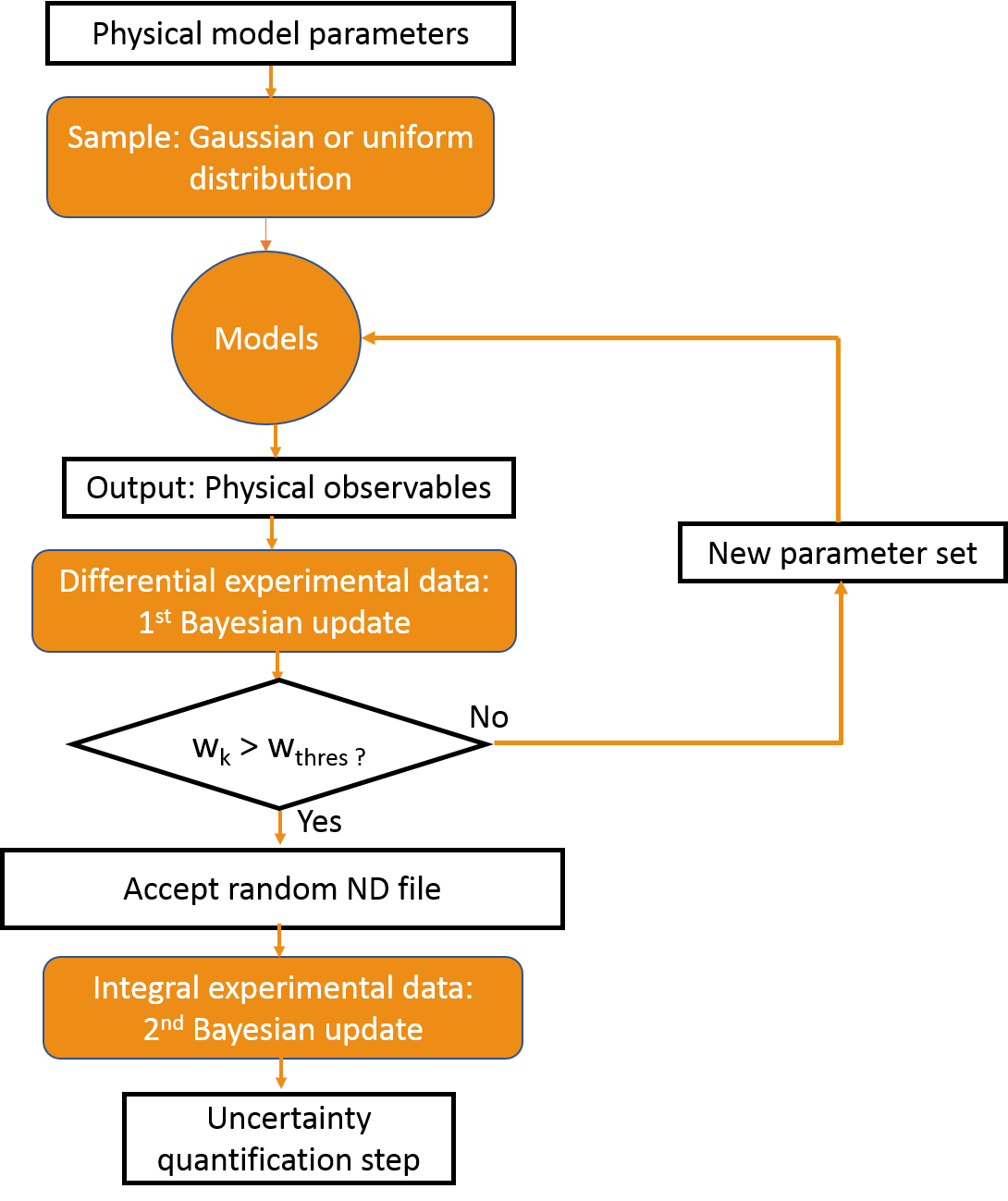} 
  \caption{A flowchart showing the Bayesian updating scheme for multi-level uncertainty propagation within the Total Monte Carlo method. $w_k$ is the weight computed for the $k^{th}$ ND file, and $w_{thres}$ is a minimum weight threshold assigned to each random ND file such that any file that does not meet this criterion is discarded. A value of $w_{thres}$ =  2.06e-09 which correspond to a total critical $\chi^2$ value of 40 was assigned to each ND file. The critical $\chi^2$ value correspond to the upper-tail critical value of the $\chi^2$ distribution with 28 degrees of freedom (varied parameters) at 95\% confidence level assuming that the $\chi^2$ distribution is in the form of a gamma distribution with the scale parameter equal to 2 and the shape parameter equal to N/2; where N is the number of the degrees of freedom. \textcolor{red}{It should be noted that, this could introduce some bias into the calculation depending on the choice of the cut-off parameter.}}
  \label{fig_BayesUpdate}
\end{figure}

In Fig.~\ref{fig_BayesUpdate}, a diagram showing the proposed Bayesian updating scheme for multi-level uncertainty propagation within the TMC method is presented. \textcolor{blue}{The term uncertainty as used in this work is represented by an estimated one standard deviation of the distribution under consideration.} The TMC method has been presented extensively in several references:~\cite{Koning-2008, Rochman-2012evaluation,koning-2015bayesianfull,Rochman-2016linearperturb,Rochman-2009uncertainties} as well as applied to many applications:~\cite{Alhassan-2013a,Alhassan-2014ANE,Alhassan-2013b,Helgesson-2013,Sjostrand-2013a,Sjostrand-2013neudos,Rochman-2012NEA,Rochman-2011SFR,Rochman-2012pwr}. 
The first step in our updating scheme involves the identification of model input parameters (and the corresponding parameter widths) that play significant roles in defining or characterizing the reaction observables of interest. These parameters were determined through a combined use of expert judgement and sensitivity analysis. The next step is to determine the distribution for each model parameter of interest. Since our goal is to use non-informative (also known as 'objective') priors as much as possible, we assumed here that, we have no prior knowledge of the parameters under consideration. Therefore the uniform distribution was used as the prior distribution for all selected model parameters. \textcolor{blue}{Also, the parameters were sampled in a non correlated way.} A complete objective prior in our case as observed also in Refs.\cite{Helgesson-2017combining,koning-2015bayesianfull}, is however not possible since the evaluation in this work made use of a nominal file which was selected by comparing TALYS results with available differential experimental data. \textcolor{blue}{The nominal file is a complete evaluation with its corresponding input parameter set and represents the evaluator's best effort before adjustment or assimilation.}

The third step is to determine the nominal file with its parameter set for adjustment. A good starting point is to use reference input parameters from the RIPL database~\cite{capote-2009RIPL} for model calculations, if available. In this work however, the nominal file (with its parameter set) was taken from the TENDL-2015 evaluation~\cite{Tendl-2015}. It should be noted that, no significant changes have been made to the TENDL-2015 evaluation of $^{208}$Pb in the recent TENDL library release~\cite{Koning-2019tendl}. Next, all the model parameters were randomly varied within their assigned parameter widths using the TALYS code system (T6)~\cite{Koning-2012b} and \textcolor{blue}{a total of  2700 random nuclear data files were produced.} 

The fourth step involves carrying out a probabilistic data adjustment or assimilation using a Bayesian updating scheme (more details are presented in section~\ref{Bayesian_update}). Two Bayesian updates are performed in this work: (1) using only differential experimental data (1st update), and (2) using integral experimental data (2nd update). The main objective of this work is to combine the individual likelihood functions from each update into a global (combined) likelihood function for each nuclear data (ND) file. From these global likelihood functions, the new 'best' file which takes both differential and integral experimental data \emph{explicitly} into account, is selected. The best file is then the file (with its parameter set) that maximises the combined (global) likelihood function. In the case of the 1st update, as shown in Fig.~\ref{fig_BayesUpdate}, \textcolor{blue}{we define an acceptance criterion such that any ND file with a weight less than an assigned threshold file weight ($w_{thres}$ = 2.06e-09) which corresponds to a critical $\chi^2$ value of 40, is discarded. The critical $\chi^2$  value correspond to the upper-tail critical value of the $\chi^2$ distribution with 28 degrees of freedom (i.e. varied model parameters; see Table~\ref{Table1modelp}) at 95\% confidence level. The critical $\chi^2$ value was used here with the assumption that the $\chi^2$ distribution (from the 1st update) is in the form of a gamma distribution with the scale parameter equal to 2 and the shape parameter equal to N/2, where N is the number of degrees of freedom. The critical $\chi^2$ values can be found in the appendix of most statistical books, for example, in Appendix C, Table G of Ref.~\cite{Bluman-2009elementary}. The relationship between the $\chi^2$ and the weight ($w_k$) is given in Eq.~\ref{weights_Eqn}. If the $\chi^2$ value (or weight) computed for a particular random ND file is greater than the critical $\chi^2$ = 40 (or lower than the corresponding threshold weight ($w_{thres}$)), it gives an indication that the file under consideration did not fit the selected experimental data and is therefore rejected.} In this way, files with very low weights were rejected which could lead to a savings in computational time. For example, in the case of this work, after setting a threshold weight (as presented in Fig.~\ref{fig_BayesUpdate}), a total of 2046 files were accepted; down from the initial 2700 files. This represents a 24\% reduction in the number of random ND files used. \textcolor{red}{It should be noted however that, this could introduce some bias into the calculation depending on the choice of the cut-off parameter. To address this, the use of Russian roulette for randomly rejecting ND files with insignificant weights was proposed in Refs.~\cite{helgesson-2014incorporating,Helgesson-2017combining}. This approach was however not used in this work}.

 \textcolor{blue}{Since the random samples were drawn from a rather large model parameter space (see section~\ref{prior_models}), a possibility could arise where a large number of the random ND files are assigned with very low or insignificant weights. Therefore an Effective Sample Size (ESS) (given by Eq.~\ref{ESS_eq}) has been computed for each update}:

\begin{equation}
ESS=\frac{\big( \sum_{k=1}^{n} w_k \big)^2}{\sum_{k=1}^{n} w_k^2} 
\label{ESS_eq}
\end{equation}

\textcolor{blue}{The ESS was used in order to determine the number of random ND files with significant weights for both the prior and posterior distributions. The ESS gives an indication on the number of random ND files with significant impact on the distributions considered. Also, in order to determine if the prior and posterior $k_{\rm eff}$ distributions have converged, convergence plots (i.e. for the mean and the ND uncertainty as a function of random samples), are presented for each update}. 


\subsection{Bayesian updating within TMC}
\label{Bayesian_update}

\subsubsection{Priors - model parameters and their distributions}
\label{prior_models}
In Table~\ref{Table1modelp}, the model parameters as well as the parameter widths (given as a fraction (\%) of their absolute values), assigned to selected model parameters are presented. The parameter widths of $a$, $g_\pi$ and $g_\nu$ as presented in the Table are given in terms of the mass number $A$. Where $a$ is the real central diffuseness, $g_\pi$ and $g_\nu$ are the single-particle state densities as used in exciton model analyses~\cite{Koning-2004global}. The parameter widths were determined by means of comparisons of scattered random TALYS curves with differential experimental data from the EXFOR database~\cite{Koning-2012b}. In order to increase the parameter space, similar to Ref.~\cite{koning-2015bayesianfull}, the model parameter widths (as presented in Table~\ref{Table1modelp}) were all multiplied by a factor of three. In addition, the parameters were sampled from uniform distributions. This was done in order to obtain non-informative priors as much as possible for each model parameter. However, as observed earlier \textcolor{blue}{in section~\ref{updating_scheme}}, a complete state of non-informative prior can never be reached since experiments were used to fine tune the default model parameters used in the TALYS code as well as in determining the nominal file. All the parameters presented were then simultaneously varied within their parameter widths in a TMC fashion to produce a total of 2700 random $^{208}$Pb nuclear data files.

 \begin{table}[htb]
       \centering
       \caption{Selected nuclear model parameters of the TALYS code (with their parameter widths) used for parameter variations. The parameter widths are given as a fraction (\%) of their absolute values. The parameter widths of $a$, $g_\pi$ and $g_\nu$ are given in terms of the mass number $A$. $a$ is the real central diffuseness, $g_\pi$ and $g_\nu$ are the single-particle state densities as used in exciton model analyses~\cite{Koning-2004global}. A complete list of all the model parameters can be found in Ref.~\cite{TALYS-2007}. \textcolor{blue}{All parameters were varied altogether within their parameter widths in a TMC fashion.}}
       \label{Table1modelp}
       \begin{tabular}{cc|cc}  
       \hline\hline
       Parameter  &  Uncertainty(\%)  & Parameter  &  Uncertainty(\%)    \\
       \hline
       $r^n_V$  &  1.5 &   $a^n_V$ &  2.0  \\
       $v^n_1$ & 1.9 & $v^n_2$  &  3.0  \\        
       $v^n_3$ &  3.1  & $v^n_4$ & 5.0  \\
       $w^n_1$ & 9.7 & $w^n_2$  &  10.0  \\ 
       $d^n_1$ &  9.4  & $d^n_2$ & 10.0  \\
       $d^n_3$  & 9.4 &  $r^n_D$ &  3.5 \\   
       $a^n_D$  & 4.0  &  $r^n_{SO}$ & 9.7 \\
       $a^n_{SO}$ & 10.0 & $v^n_{SO1}$ & 5.0 \\
       $v^n_{SO2}$ & 10.0 & $w^n_{SO1}$ &  20.0 \\
       $w^n_{SO2}$ & 20.0 & $\Gamma_\gamma$ &  5.0 \\
       $a(^{207}Pb)$ &  4.5  & $a(^{206}Pb)$ & 6.5\\
       $a(^{208}Pb)$&  5.0 & $a(^{205}Pb)$ &  6.5   \\
       $\sigma^2$  &  19.0 & $M^2$  & 21.0 \\
       $g_\pi$($^{207}Pb$) &  6.5 & $g_\nu$($^{207}Pb$) & 6.5 \\
       \hline\hline
       \end{tabular}
       \end{table} 

From Table~\ref{Table1modelp}, $r$ represents the real central radius, $V$ is the volume-central, $D$ and $SO$ respectively are the surface-central and spin-orbit potentials and $W$ is the imaginary depth of the optical model~\cite{TALYS-2007}. $\sigma^2$ is the spin cut-off parameter which represents the width of the angular momentum distribution of the level density~\cite{TALYS-2007}. The superscript $n$ denotes neutron induced reactions while the subscripts $v$ and $s$ respectively denote the real volume and real surface of the optical model. $\Gamma_{\gamma}$ is the average radiative capture width; $M^2$ is the average squared matrix element of the residual interaction as used in exciton model analyses~\cite{Koning-2004global}. The default optical model potentials (OMP) used in TALYS are the local and global parametrization of Koning and Delaroche (see Ref.~\cite{Koning-2003local}) and therefore were also used in this work.

\subsubsection{1st Bayesian update: using differential experimental data}
\label{1st_update}
Selected differential experimental data from the EXFOR database~\cite{EXFOR-2007} were used for the first Bayesian update. The selection of experimental data sets were based on a number of criteria:  \textcolor{blue}{(1) We do not consider experiments that are inconsistent with other experimental sets and also deviates from the trend of other evaluations (if available) as well as our model calculations,} (2) \textcolor{red}{We assume 10\% uncertainty for experimental data sets for which no uncertainties were reported but which however were reviewed and classified as good quality experiments in Ref.~\cite{koning-2014statistical}. As noted in Ref.~\cite{koning-2015bayesianfull}, these classifications are usually subjective in nature and could therefore introduce some bias into the evaluation process. However, (as observed also in Ref.~\cite{koning-2015bayesianfull}), blindly using all experiments in EXFOR without selection (or assigning weights) usually leads to very large $\chi^2$ values between model calculations and experiments.} 3) If a particular author repeats a set of experiments for the same or similar energy range for the same cross section (and they are found to be inconsistent), we select the updated (or current) measurements, (4) In some cases, we also drop the first and/or last measured points \textcolor{blue}{(with respect to incident neutron energy)} since these measurements can easily be dominated with background noise. 

In Table~\ref{Exp_data}, we present the experimental data that were selected for the 1st update showing the number of data points per reaction channel, the corresponding MT numbers (\textcolor{blue}{in ENDF terminology)}, the EXFOR ID, the first author and the year the measurements were carried out. It should be noted that only data points that fall between 5 and 20 MeV were used for the adjustments and therefore presented.

 \begin{table}[h!]
       \centering
       \caption{\textcolor{blue}{Differential experimental data used for the adjustment of $^{208}$Pb showing the number of data points per reaction channel used. Only data points between 5 and 20 MeV were used in the computation of file weights. Also, only the names of the first authors have been presented.}}
       \label{Exp_data}
       \begin{tabular}{cccccc}  
       \hline\hline
       \pbox{20cm}{Cross \\ section}  & \pbox{20cm}{ MT \\ number}  & \pbox{20cm}{Data points used \\ 5 - 20 MeV}   & Author  & EXFOR ID  & Year \\
       \hline
       (n,tot) & 001 & 142   &  R.F. Carlton  &  13735002 & 1991  \\        
       (n,tot) & 001  &  116  & D.G. Foster Jr & 10047092 & 1971  \\
       (n,non-el) & 003 &  1 & G.M.Haas & 11794005 & 1963 \\
       (n,inl)  & 004  & 46   & L.C. Mihailescu & 23039003 & 2008 \\
       (n,2n) & 016 & 16 & Frehaut & 20416058 & 1980 \\
       (n,$\gamma$) &  102  &  8   &  I. Bergqvist &  10226002 &  1972 \\
       (n,$\gamma$) &  102 &  7   & J. Csikai  &  30074007 &  1967 \\
        (n,$\gamma$) &  102 &  1   & D. Drake &   10193006 &  1971 \\
       \hline\hline
       \end{tabular}
       \end{table} 

For the 1st Bayesian update we consider the following: given that, the probability distribution of the model parameters ($p$) before any data (cross sections in this case) were observed (also known here simply as the prior) is $P^{prior}(\sigma_T,p)$, and $L(\sigma_E|\sigma_T,p)$ is the probability of the experimental data occurring given that the model parameters are accurate (also known as the likelihood function); the posterior, which is the probability of the parameter values being accurate given differential experimental data ($P^{post}(\sigma_T,p|\sigma_E)$), can be given as:

\begin{equation}
P^{post}(\sigma_T,p|\sigma_E) \propto L(\sigma_E|\sigma_T,p) \times P^{prior}(\sigma_T,p)
\end{equation}

where $\sigma_T$ and $\sigma_E$ denote the TALYS calculated and experimental observables (cross sections in this case) respectively. \textcolor{blue}{If we assume that the off-diagonal elements of the experimental covariance matrix is zero, we can compute a reduced chi square ($\chi^2_{c(k)}$) per reaction channel and for the random ND file $k$, similar to Ref.~\cite{koning-2015bayesianfull} as follows:}

\begin{equation}
    \chi^2_{c(k)} = \frac{1}{N_p} \sum_{i=1}^{N_p} \bigg(  \frac{\sigma^i_{T(k)} - \sigma^i_E}{\Delta \sigma^i_E} \bigg)^2
    \label{gen_chi2}
\end{equation}

\textcolor{blue}{where $\sigma^i_{T(k)}$ is a vector of TALYS calculated observables as a function of incident neutron energy ($i$), found in the $k^{th}$ random ND file for the channel $c$ and $\sigma^i_E$ is a vector of experimental observables for the channel $c$ at incident neutron energy ($i$),  $\Delta \sigma^i_E$ is the experimental uncertainty at an incident energy $i$ of channel $c$, and $N_p$ is the total number of experimental points per reaction channel considered. Where there were no matches in energy ($i$) between TALYS output and the experimental data set for the $c^{th}$ channel, \textcolor{red}{TALYS data were interpolated to match corresponding experimental values. For the purpose of this work, the reduced $\chi^2$ and the $\chi^2$ are used interchangeably and refer to the same thing.}}

 \textcolor{red}{One major difference in the computation of the chi square presented in Eq.~\ref{gen_chi2} and that used in Ref.~\cite{koning-2015bayesianfull} is that, while the average is taken per experimental data set in Ref.~\cite{koning-2015bayesianfull}, the average is simply taken per channel in this work. By averaging per channel, we assign all experimental data points per channel with equal weights. Also, in order to give each channel equal weight, we then take an average over the considered channels. Additionally, in Ref.~\cite{koning-2015bayesianfull}, the effects of model defects were taken into account by normalizing the $\chi^2_{E(k)}$ for each random ND file (as presented in Eq.~\ref{Likelihoodfun1}) with the chi square obtained for the global TALYS calculation (usually TALYS run 0). A similar solution was presented also in Ref.~\cite{Bauge-2011} where instead of normalizing the  $\chi^2_{E(k)}$ with the global TALYS calculation, the $\chi^2_{E(k)}$ was normalized with the minimum chi square ($\chi^2_{min}$). Since the underlying assumption for our Bayesian update is that the models used are 'good enough', no modifications as in Refs.~\cite{koning-2015bayesianfull,Bauge-2011}, were made to the likelihood function. Since our models are not perfect, this assumption is entirely not true, however, we think that the quality of the models used are adequate for the purpose of this evaluation.} 
 
From Eq.~\ref{gen_chi2}, we can compute a weighted chi square ($\chi_{E(k)}^2$) given by:
\begin{equation}
\chi_{E(k)}^2 = \frac{\sum_{c=1}^{N_c} w_c \chi^2_{c(k)}}{\sum_{c=1}^{N_c} w_c}
\label{weighted_chi2}
\end{equation}

where $w_c$ is the channel weight and $N_c$ is the number of considered channels. \textcolor{red}{Two types of channel weights were used in this work: (1) $w_c = \overline{\sigma_c}$ and (2) $w_c = 1/N_c$ (also called unweighted channels since all channels have equal weights).  $\overline{\sigma_c}$ is the average cross section for the channel $c$. By assigning channel weights equal to the average cross section as in (1), we give more weight to channels with relatively large cross sections such as the (n,tot), (n,inl) and the (n,2n) cross sections of  $^{208}$Pb while channels such as the (n,$\gamma$) with smaller average cross sections are assigned with low channel weights. This is however not consistent for a 'general purpose' library since the fit would favour some cross sections than others. In (2), all considered channels are assigned with equal weights and therefore all channels are of equal importance in the adjustment procedure. Case (2) is therefore referred to as the unweighted channel case in this work. From the $\chi_{E(k)}^2$ computed, the likelihood function can now be computed using an expression given within the Unified Monte Carlo approach (UMC-B)~\cite{Capote-2012UMC-B} and utilized also in the TMC + UMC-B method presented in Ref.~\cite{Helgesson-2017combining}:}

\begin{equation}
  L(\sigma_E|\sigma_T,p) \propto exp \left(-\frac{\chi^2_{E(k)}}{2} \right)
  \label{Likelihoodfun1}
\end{equation}

The motivation for using Eq.~\ref{Likelihoodfun1} has been presented previously in Ref.~\cite{Helgesson-2017combining}. From Eq.~\ref{Likelihoodfun1}, each random ND file in the case of the 1st update was assigned a file weight based on only selected differential experimental data from EXFOR using Eq.~\ref{weights_Eqn}:
\begin{equation}
w_{E(k)} = e^{-\frac{1}{2}\chi^2_{E(k)}}
\label{weights_Eqn}
\end{equation}

where $w_{E(k)}$ is the weight assigned to the $k^{th}$ ND file. In order to relate the file weight values to one, the weights were normalized with their maximum weight as follows:
\begin{equation}
w_{E(k),N} = \frac{w_{E(k)}}{max(w_{E(k)})}
\label{weights_Eq1}
\end{equation}

where $max(w_{E(k)})$ and $w_{E(k),N}$ are the maximum and the normalized weights from the 1st update respectively. The weights computed for the 'best' files from this work have been compared with weights computed for the ENDF/B-VIII.0, JEFF-3.3, JENDL-4.0, TENDL-2017 and CENDL-3.1 nuclear data libraries using the same methodology and experimental data and presented in Table~\ref{compare_weights11}.

\subsubsection{2nd Bayesian update: using integral experimental data}
In Table~\ref{table_critB}, the benchmark experiments used for the 2nd update are presented showing the experimental benchmark $k_{\rm eff}$ and the experimental benchmark uncertainty. These benchmarks can be obtained from the International Handbook of Evaluated Criticality Safety Benchmark Experiments (ICSBEP)~\cite{Briggs-2003crit}. From the Table, HMF stands for Highly Enriched Uranium (HEU) Metallic Fast, PMF for Plutonium Metallic Fast, while LCT stands for Low Enriched Uranium (LEU) Compound Thermal benchmarks. 
 \begin{table}[h!]
  \begin{center}
  \footnotesize
  \centering
  \tabcolsep=0.10cm
    \caption{\label{table_critB} The integral experiments used in this work showing the evaluated experimental benchmark $k_{\rm eff}$ and corresponding experimental uncertainties. These benchmarks were obtained from the (ICSBEP) handbook~\cite{Briggs-2003crit}. HMF stands for Highly Enriched Uranium (HEU) Metallic Fast, PMF for Plutonium Metallic Fast and LCT for Low Enriched Uranium (LEU) Compound Thermal benchmarks.} 
    \begin{tabular}{lcccc}
    \toprule
    Benchmark  category  & Abbreviation & \pbox{20cm}{Experimental \\ Benchmark $k_{\rm eff}$} & \pbox{20cm}{Experimental Benchmark \\ uncertainty (pcm)}  \\
    \midrule
    PMF035 case 1  & pmf35c1 & 1.0000 & 160 \\
    HMF027 case 1  & hmf27c1 & 1.0000 & 250  \\
    HMF057 case 1  & hmf57c1 & 1.0000 & 200  \\
    HMF057 case 2  & hmf57c2 & 1.0000 & 230  \\
    HMF057 case 3  & hmf57c3 & 1.0000 & 320  \\
    HMF057 case 4  & hmf57c4 & 1.0000 & 400  \\
    HMF057 case 5  & hmf57c5 & 1.0000 & 190  \\
    HMF064 case 1  & hmf64c1 & 0.9996 & 80   \\
    LCT010 case 1  & lct10c1 & 1.0000  & 210  \\
    LCT010 case 20  & lct10c20 & 1.0000  & 280  \\
    \bottomrule
    \end{tabular}
  \end{center}
\end{table}

Given that $P^{prior}(k^B_{\rm eff,cal}|\sigma_T)$ is the probability distribution of the $k_{\rm eff}$ given cross sections from the 1st update (i.e. before the inclusion of integral benchmark data) and $L(k^B_{\rm eff,exp}|k^B_{\rm eff,cal},\sigma_T)$ is our likelihood function; the posterior distribution $(P^{post}(k^B_{\rm eff,cal},\sigma_T|k^B_{\rm eff,exp})$, can be expressed as:

\begin{equation}
P^{post}(k^B_{\rm eff,cal},\sigma_T| k^B_{\rm eff,exp}) \propto L(k^B_{\rm eff,exp}|k^B_{\rm eff,cal},\sigma_T) \times P^{prior}(k^B_{\rm eff,cal}|\sigma_T)
\label{2nd_BayesEq}
\end{equation}

Where the likelihood function is given by:

\begin{equation}
    L(k^B_{\rm eff,exp}|k^B_{\rm eff,cal},\sigma_T) \propto exp \left(-\frac{\chi^2_{B(k,j)}}{2}\right)
    \label{weights_general}
\end{equation}

\textcolor{blue}{In the case of one benchmark (as utilized in this work)}, the chi square ($\chi^2_{B(k,j)}$) as a function of random ND ($k$) and benchmark $j$, can be expressed as:

\begin{equation}
    \chi^2_{B(k,j)} =\Bigg( \frac{k^B_{\rm eff(k,j)} - k^B_{\rm eff,E(j)}}{\Delta k^B_{\rm eff,E(j)}} \Bigg)^2
    \label{Eq_simpleChi2}
\end{equation}

where $k^B_{\rm eff(k,j)}$ is a vector of calculated response parameters for the $k^{th}$ random nuclear data file and the $j^{th}$ benchmark, $k^B_{\rm eff,E(j)}$ is a vector of integral benchmark experimental observables with corresponding experimental uncertainty ($\Delta k^B_{\rm eff,E(j)}$) for the $j^{th}$ benchmark. Since only one benchmark was used in the adjustment, no correlations between benchmarks were considered. The $k^B_{\rm eff(k,j)}$ in Eq.~\ref{Eq_simpleChi2} was computed by varying $^{208}$Pb nuclear data while holding constant, all other isotopes contained in the benchmark under consideration. \textcolor{red}{In the computation of the $\chi^2$ as presented in Eq.~\ref{Eq_simpleChi2}, only benchmark experimental uncertainties were taken into account. However, as noted earlier in Ref.~\cite{Alhassan-2015NDreduction}, because computer codes such as MCNP~\cite{Briesmeister-2000} are normally used for benchmark calculations and analyses, these benchmarks also contain calculation uncertainties. These uncertainties could come from (for example), our inability to model the benchmark geometry accurately with the code used, statistics (i.e. in the case where a Monte Carlo code is used) or from uncertainties due to nuclear data (since nuclear data are used as inputs to these codes). These calculation uncertainties were however not considered in this work.}

\textcolor{blue}{Using the $\chi^2_{B(k,j)}$ presented in Eq.~\ref{Eq_simpleChi2}, weights similar to Eq.~\ref{weights_Eqn}, were assigned to each random nuclear data based on experimental data from the hmf57c1 benchmark. The weights were also normalized with the maximum weight in order to keep the weight values between 0 and 1. As a proof of concept, we have only used one benchmark - the  hmf57c1 benchmark for adjustment in this work; all other benchmarks (as presented in the Table~\ref{table_critB}) were used for testing the adjusted file's performance. The motivation to use the hmf57c1 is because, aside being a fast neutron spectrum benchmark, it  was observed to be sensitive to $^{208}$Pb nuclear data. Additionally, it has a relatively small benchmark uncertainty of 200 pcm and also comes with five different benchmark cases which can be used for verification purposes. A benchmark case as used in this work (as also used in Ref.~\cite{Briggs-2003crit}), is used to define a series of similar benchmarks where a limited number of parameters (e.g. geometrical parameters) have been varied.}
 
After incorporating the weights in the 2nd Bayesian update, an updated (or weighted) mean and the corresponding weighted variance were computed for each random ND file using Eq.~\ref{weighted_mean} and ~\ref{variance_1}: 

\begin{equation}
    \overline{k^B_{\rm eff(k,w)}} = \frac{\sum\limits_{k=1}^{n} w_{B(k,j)} \cdot k^B_{\rm eff(k,j)}}{\sum\limits_{k=1}^{n} w_{B(k,j)}}
    \label{weighted_mean}
\end{equation}

\begin{equation}
    Var(\overline{k^B_{\rm eff(k,w)})} = \frac{\sum\limits_{k=1}^{n} w_{B(k,j)} k^B_{\rm eff(k,j)}}{\sum\limits_{k=1}^{n} w_{B(k,j)}} - \overline{k^B_{\rm eff(k,w)}}^2
    \label{variance_1}
\end{equation}

where $\overline{k^B_{\rm eff(k,w)}}$ and $Var(\overline{k^B_{\rm eff(k,w)}})$ are the weighted mean and weighted variance respectively, and $w_{B(k,j)}$ is the file weight assigned to the $k^{th}$ random ND file using experimental information from the benchmark $j$.

\subsubsection{Global likelihood function: Combining weights from differential and integral experiments}
According to Ref.~\cite{lista-2017combination}, a statistically rigorous way to combine experimental data from different measurements is to combine the likelihood functions computed for the individual measurements. Similar to ~\cite{lista-2017combination}, we combine here the likelihood functions computed from the 1st and 2nd updates into a global likelihood function for each random ND file. From probability theory, if two events are independent, the probability of both events occurring is given as the product of the probabilities of each event occurring. The likelihood function in this case can be considered as the probability of the experimental data occurring given that model parameters are accurate. Therefore, if we assume that the differential and integral experimental data are independent or uncorrelated, a global (combined) likelihood function (also known here as the combined weight) can be \textcolor{blue}{obtained} as a product of the individual likelihood functions computed from differential experimental data ($L(\sigma_E|\sigma_T,p)$), and integral benchmark data ($L(k^B_{\rm eff,exp}|k^B_{\rm eff,cal},\sigma_T)$):
\begin{equation}
    L(\sigma_E,k^B_{\rm eff,exp}|\sigma_T,p,k^B_{\rm eff,cal}) = L(\sigma_E|\sigma_T,p) L(k^B_{\rm eff,exp}|k^B_{\rm eff,cal},\sigma_T) 
    \label{global_liklihood}
\end{equation}

where $L(\sigma_E,k^B_{\rm eff,exp}|\sigma_T,p,k^B_{\rm eff,cal})$ is the global (combined) likelihood function. From Bayes theorem, our new posterior distribution ($P^{post}_{comb} (k^B_{\rm eff,cal}|\sigma_E,k^B_{\rm eff,exp}$)) which takes into consideration information from both differential and integral benchmark experiments, can be given as: 
\begin{equation}
P^{post}_{comb} (k^B_{\rm eff,cal}|\sigma_E,k^B_{\rm eff,exp}) = P^{prior}(k^B_{\rm eff,cal}|\sigma_T) L(\sigma_E|\sigma_T,p) L(k^B_{\rm eff,exp}|k^B_{\rm eff,cal},\sigma_T) 
\label{comb_BayesEq}
\end{equation}

From Eq.~\ref{global_liklihood}, a combined weight ($w_{T,k}$) which is equal to the global likelihood function, is computed for each $k$ ND file: 
\begin{equation}
w_{T,k} = w_{E(k)} \cdot w_{B(k,j)}
\label{global_weight}
\end{equation}
where $w_{E(k)}$ and $w_{B(k,j)}$ are the weights for the $k^{th}$ random ND file computed using differential experimental data and integral benchmark experiment $j$ respectively. In order to compare $w_{E(k)}$ and $w_{B(k,j)}$, the weights computed for each update were both normalized using their maximum weights. In this way, the combined weight was constrained between 0 and 1. An algorithm was developed to automatically select the new 'best' file i.e. the file with the parameter set that maximizes the global (combined) likelihood function. The performance of the 'best' file has been compared against differential data from EXFOR and evaluations from the major nuclear data libraries using the reduced $\chi^2$ as the goodness of fit estimator and presented in Table~\ref{compare_weights11} as well as with integral benchmarks (see Table~\ref{Exp_data_Newresults}).

\subsection{Simulation and  analysis}
Since resonance structure extends well beyond 2 MeV for the (n,tot) and the (n,el) cross sections of $^{208}$Pb, the neutron energy range considered in the 1st update (using differential data only) was from 5 to 20 MeV. However, in order to create a complete ENDF file from 0 to 20 MeV as normally required for applications, \textcolor{red}{the entire MF-2 (in ENDF terminology for the resonance parameters) was adopted from the JEFF-3.1~\cite{Koning-2006Jeff} library} using the TARES code~\cite{TARES-2011}, while the elastic angular distributions (MF4-MT2) were adopted from ENDF/B-VII.1~\cite{Chadwick-2011} evaluation. The nuclear reaction code TALYS-1.6~\cite{TALYS-2007} was used for the computation of all reaction observables while the TEFAL code~\cite{TEFAL-2010} was used for converting TALYS output into the well known ENDF format.  

The random ENDF ND files were then processed into the ACE format at 293K using NJOY99 version 336~\cite{Briggs-2003crit}. The ACE formatted random files were used in the MCNPX code (version 2.5) for the computation of the $k_{\rm eff}$. For each benchmark, criticality calculations were carried out with 5000 neutron particles for 500 criticality cycles resulting in an average statistical uncertainty of 48 pcm. In order to speed up calculation time, the fast TMC as presented in Ref.~\cite{Rochman-2013fast} and utilized in Refs.~\cite{Alhassan-2014ANE,Sjostrand-2013a} for uncertainty propagation was also used in this work. Hence, the seed of the MCNPX code was changed for each random run by means of the DBCN card.

\section{Results}
\subsection{1st Bayesian Update: EXFOR data}
\label{1st_update}
\begin{figure}[h!] 
  \centering
  \includegraphics[trim = 5mm 70mm 5mm 65mm, clip, width=0.48\textwidth]{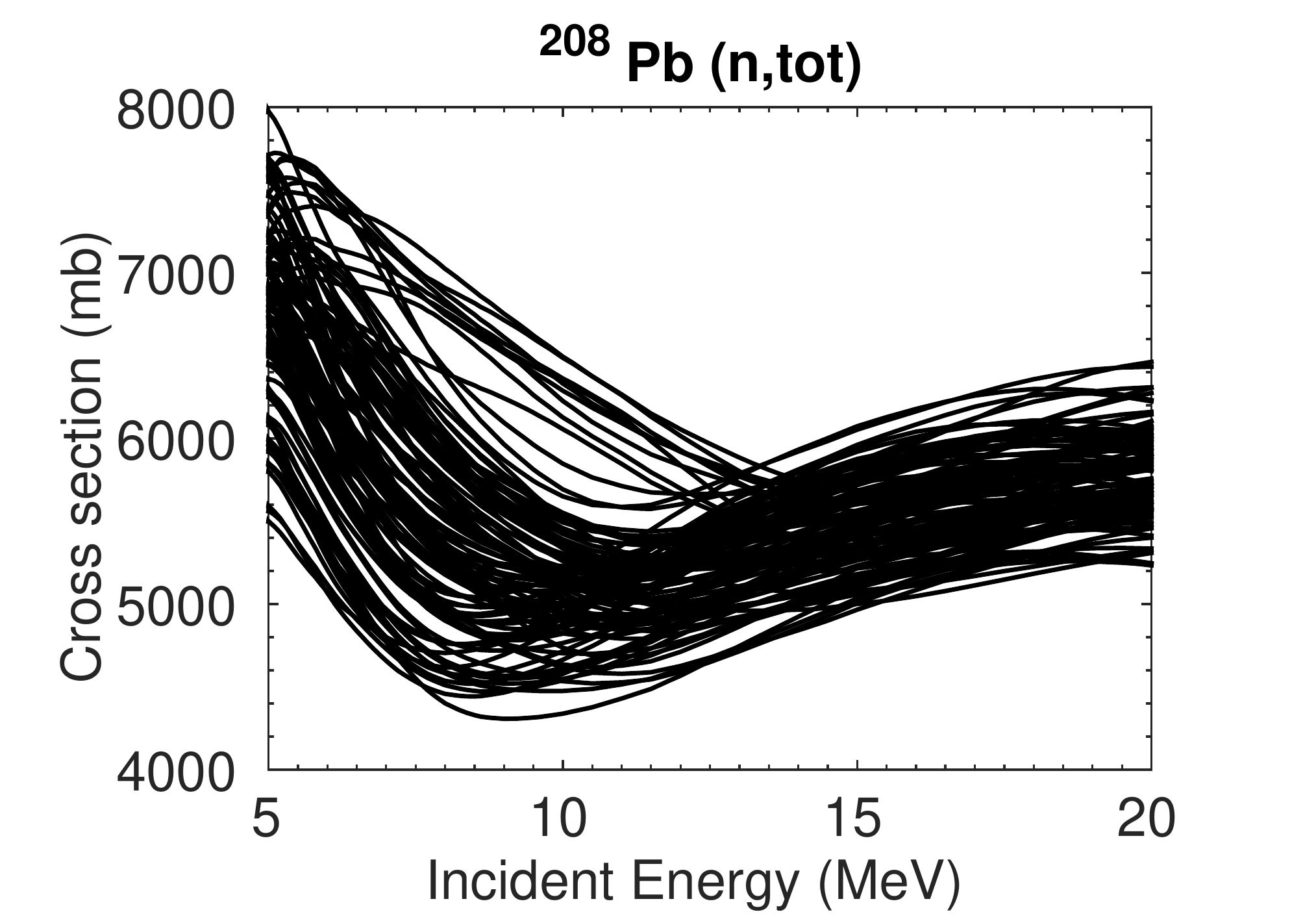} 
   \includegraphics[trim = 5mm 70mm 5mm 65mm, clip, width=0.48\textwidth]{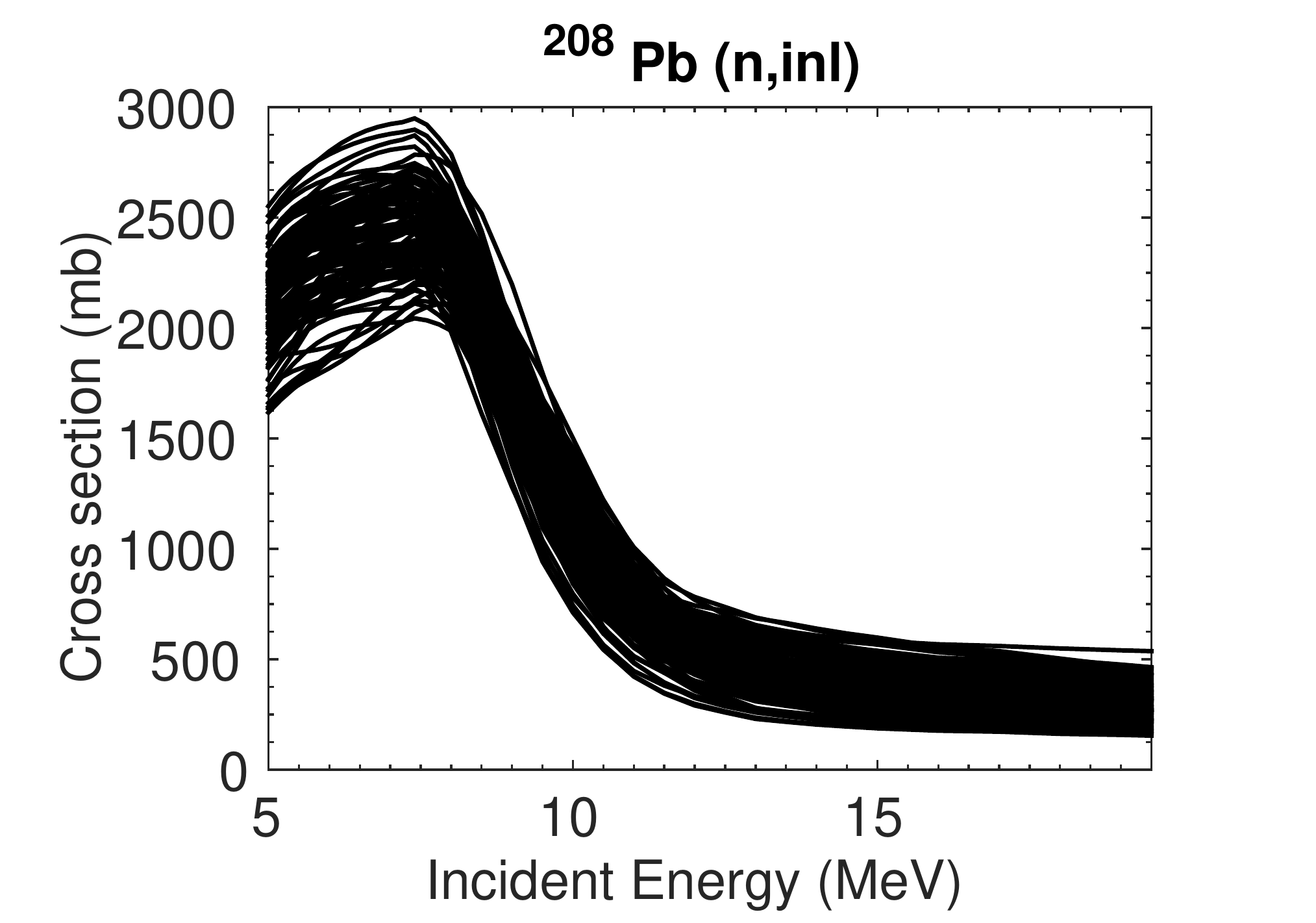} 
  \caption{Distributions of (n,tot) and (n,inl) $^{208}$Pb cross sections for 100 random cross sections. These random cross sections are used as prior for the 2nd Bayesian update.}
  \label{Distr_of_XS}
  \end{figure}

The distributions of cross sections as a function of incident neutron energy for the $(n,tot)$ and $(n,inl)$ cross sections are presented for 100 random $^{208}$Pb cross sections in Fig.~\ref{Distr_of_XS}. The spread observed in the cross sections  can be attributed to the variation of model parameters using the TMC method. These cross sections, were used as the prior for the second Bayesian update.
%

\begin{figure}[h!] 
  \centering
  \includegraphics[trim = 5mm 70mm 5mm 65mm, clip, width=0.47\textwidth]{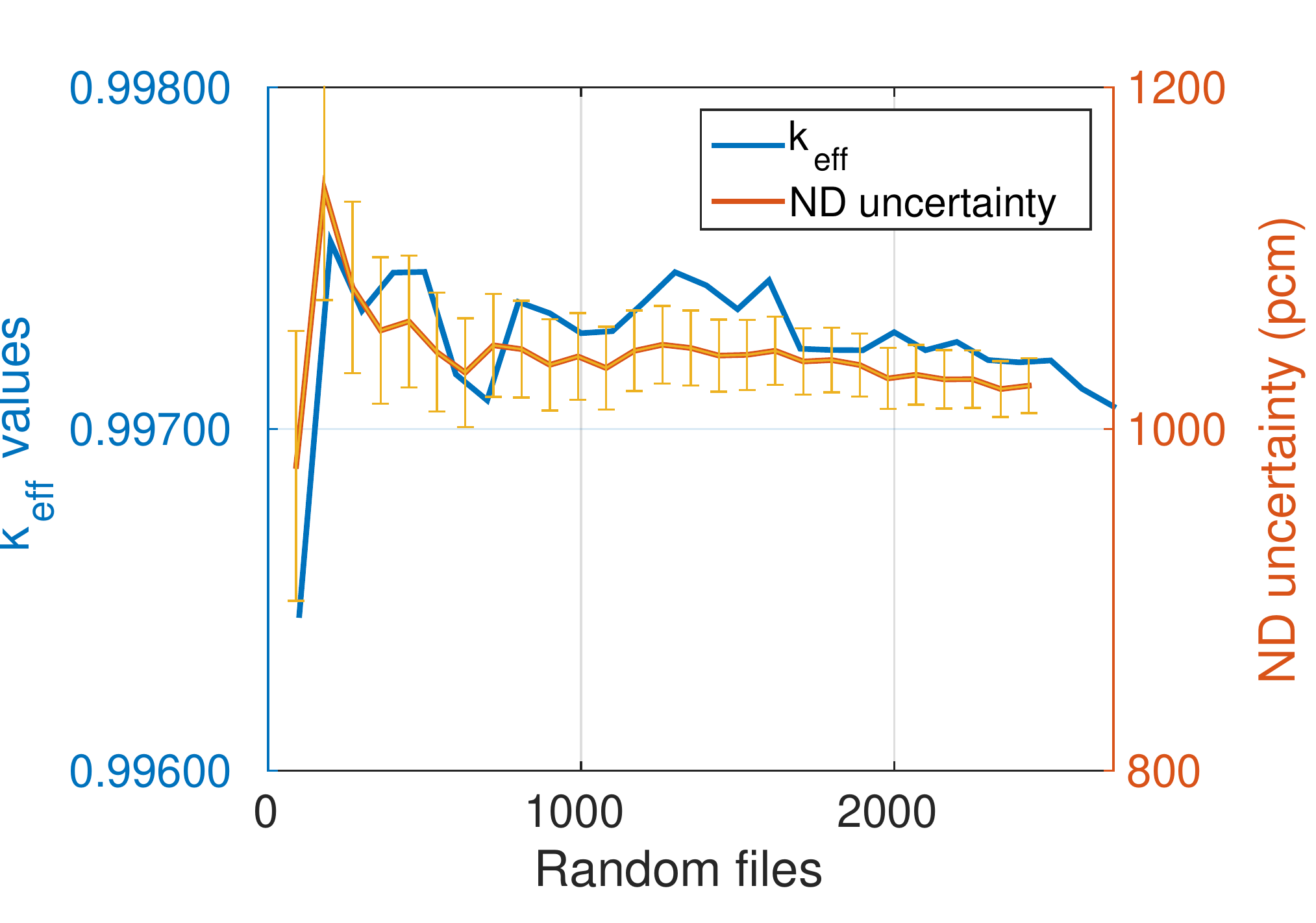} 
   \includegraphics[trim = 5mm 70mm 5mm 65mm, clip, width=0.47\textwidth]{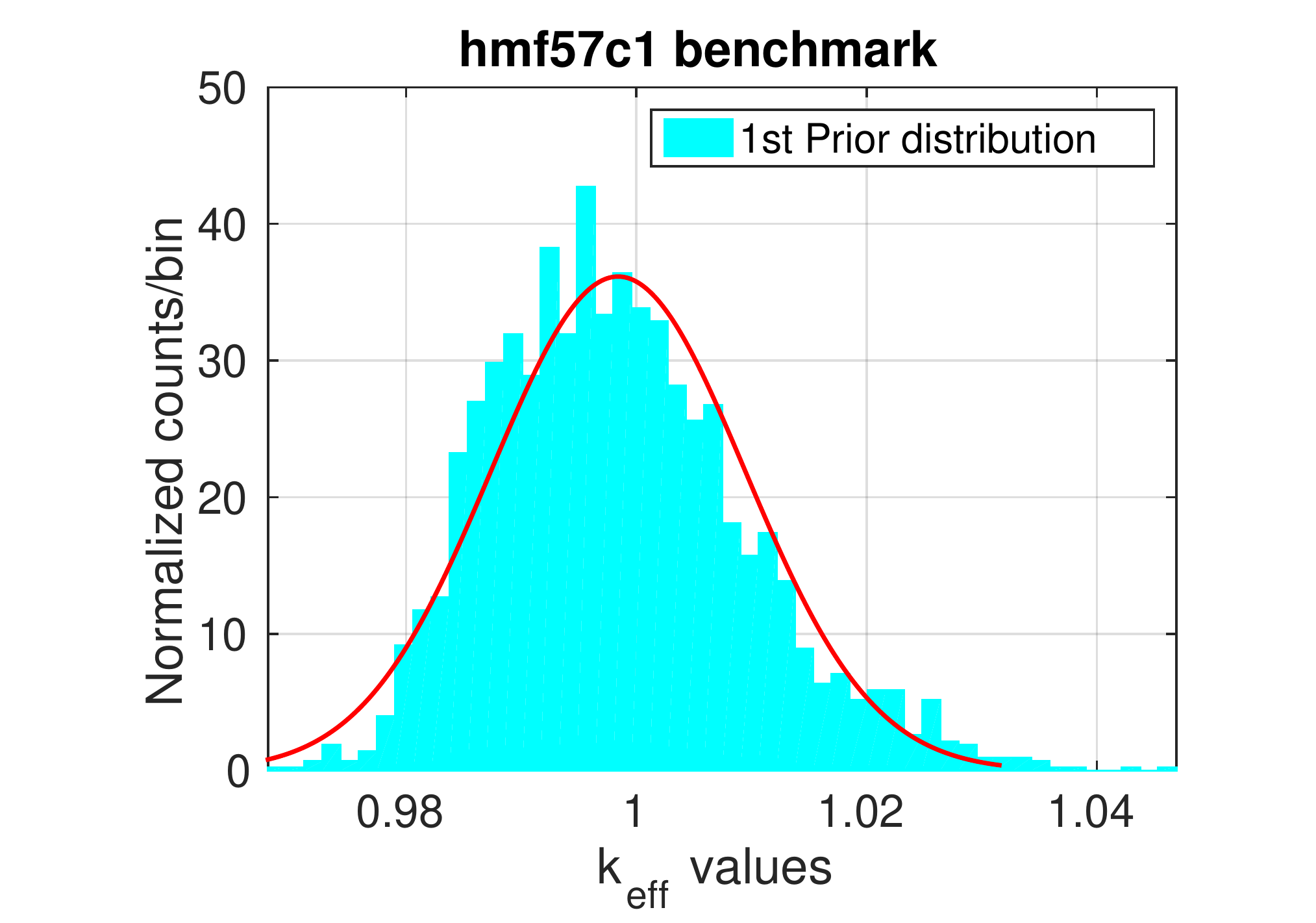} 
  \caption{\textcolor{blue}{Convergence in the mean and the $^{208}$Pb ND uncertainty \textcolor{red}{(left)} with the corresponding prior $k_{\rm eff}$ distribution \textcolor{red}{(right). The $k_{\rm eff}$ distribution is the 1st prior distribution obtained by varying $^{208}$Pb nuclear data in the hmf57c1 benchmark (a total of 2700 random ND files were used). By 1st prior distribution, we refer to the distribution of the $k_{\rm eff}$ before the implementation of the threshold weight presented in Fig.~\ref{fig_BayesUpdate}}. The error bars on the ND uncertainty represent the estimated uncertainty on the ND data uncertainty obtained for each iteration. More details on how to compute this uncertainty has been presented in Refs.~\cite{Alhassan-2014ANE,Helgesson-2013,Sjostrand-2013a}. \textcolor{blue}{The $k_{\rm eff}$ distribution is fitted to a normal distribution for the purpose of eye guidance only}.}}
  \label{cov_moments}
  \end{figure}

\textcolor{blue}{In order to determine if the random nuclear data files for the prior $k_{\rm eff}$ distribution converged, the mean (of the prior $k_{\rm eff}$ distribution) with its corresponding ND uncertainty is plotted against the number of random files for a number of iterations and presented in Fig.~\ref{cov_moments}. The $k_{\rm eff}$ distribution which is the 1st prior distribution obtained by varying $^{208}$Pb nuclear data in the hmf57c1 benchmark is also presented (\textcolor{red}{right of Fig.~\ref{cov_moments}). By 1st prior distribution, we refer to the $k_{\rm eff}$ distribution before the implementation of the threshold weight criterion presented in Fig.~\ref{fig_BayesUpdate}}. It should be noted that, the $k_{\rm eff}$ distribution presented in Fig.~\ref{cov_moments} (as in the case for all other $k_{\rm eff}$ distributions in this work), \textcolor{blue}{is fitted to a normal distribution for the purpose of eye guidance only}. \textcolor{red}{From left of the figure (Fig.~\ref{cov_moments}),} even though some fluctuations can be observed in the convergence plots shown, the percentage change in the last five iterations (for both the mean and the ND uncertainty) was observed to be less than 1\%. The error bars on the ND uncertainty represent the estimated uncertainty on the ND data uncertainty obtained for each iteration. More details on how to compute the uncertainty on the ND uncertainty was presented in Refs.~\cite{Alhassan-2014ANE,Helgesson-2013,Sjostrand-2013a}. It can be observed from the figure that, the uncertainty on the ND uncertainty estimated decreases with increasing sample size as expected.}

From the $k_{\rm eff}$ distribution presented (right of Fig.~\ref{cov_moments}), a mean, standard deviation and skewness of 0.99841, 1104 pcm and 0.55 respectively, were obtained. The positively (slightly) skewed $k_{\rm eff}$ distribution is in agreement with earlier results obtained for $^{208}$Pb nuclear data variation in fast criticality systems~\cite{Rochman-2009} and for the European Lead Cooled Training Reactor (ELECTRA)~\cite{Alhassan-2014ANE}. The skewness observed is due to the relatively high calculated $k_{\rm eff}$ values obtained for the hmf57c1 benchmark. According to Ref.~\cite{Briggs-2003crit}, these high $k_{\rm eff}$ values give an indication that the current lead cross sections have some inaccuracies which could result in more neutron reflection than is warranted by the experimental results.

\textcolor{blue}{In this work, two types of computations of the posterior distributions have been performed: (1) using weighted channels and (2) unweighted channels. As mentioned earlier in section~\ref{1st_update}, in the case of weighted channels, each channel was assigned a weight equal to its average cross section over the energy range of interest while the unweighted channels represents the case where all channels were assigned equal weights. It should be noted however that, \emph{only the plots} for the weighted case have been presented through out this work.}

\begin{figure}[h!] 
  \centering
     \includegraphics[trim = 5mm 70mm 5mm 60mm, clip, width=0.48\textwidth]{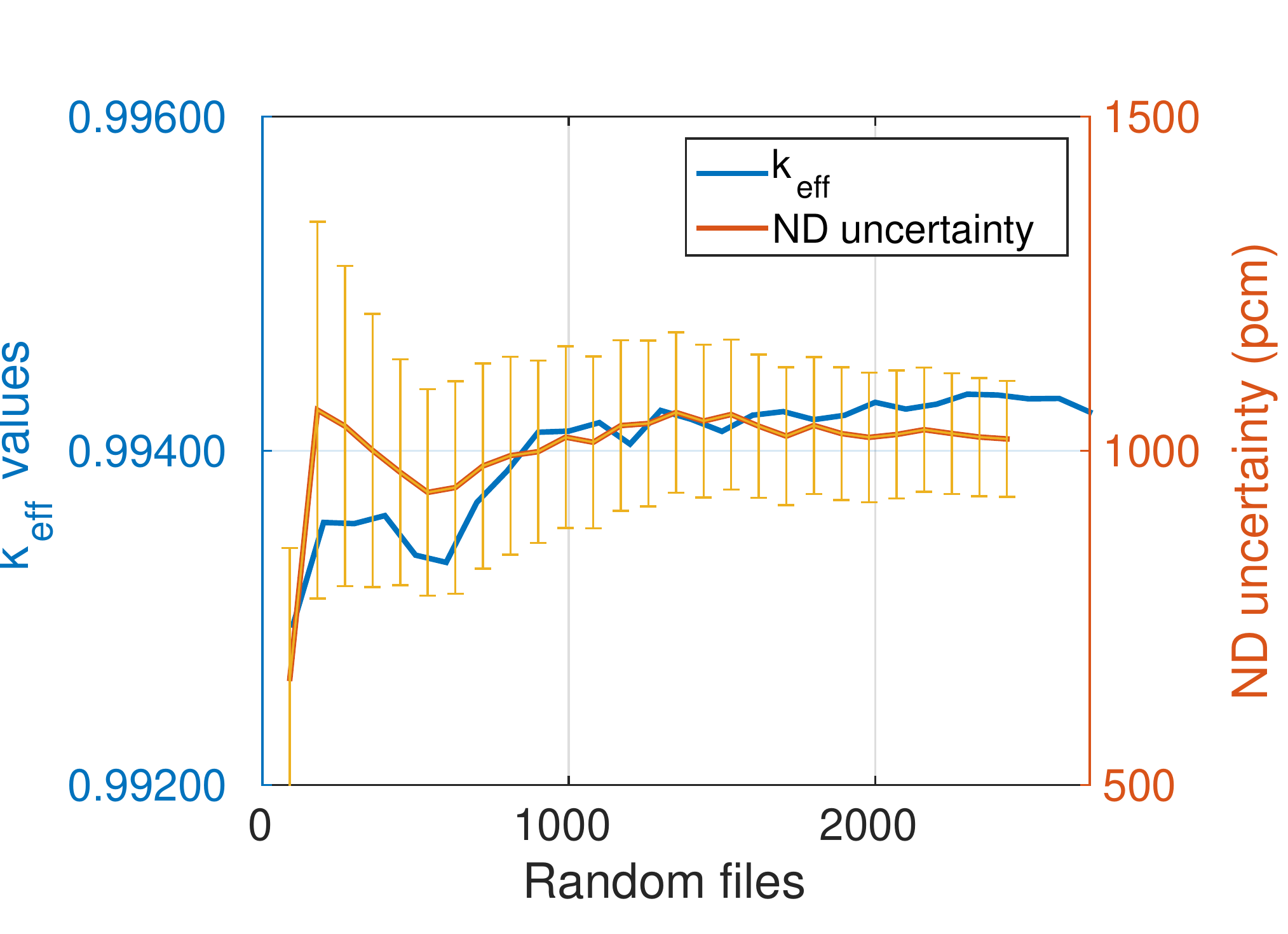} 
   \includegraphics[trim = 5mm 70mm 5mm 60mm, clip, width=0.48\textwidth]{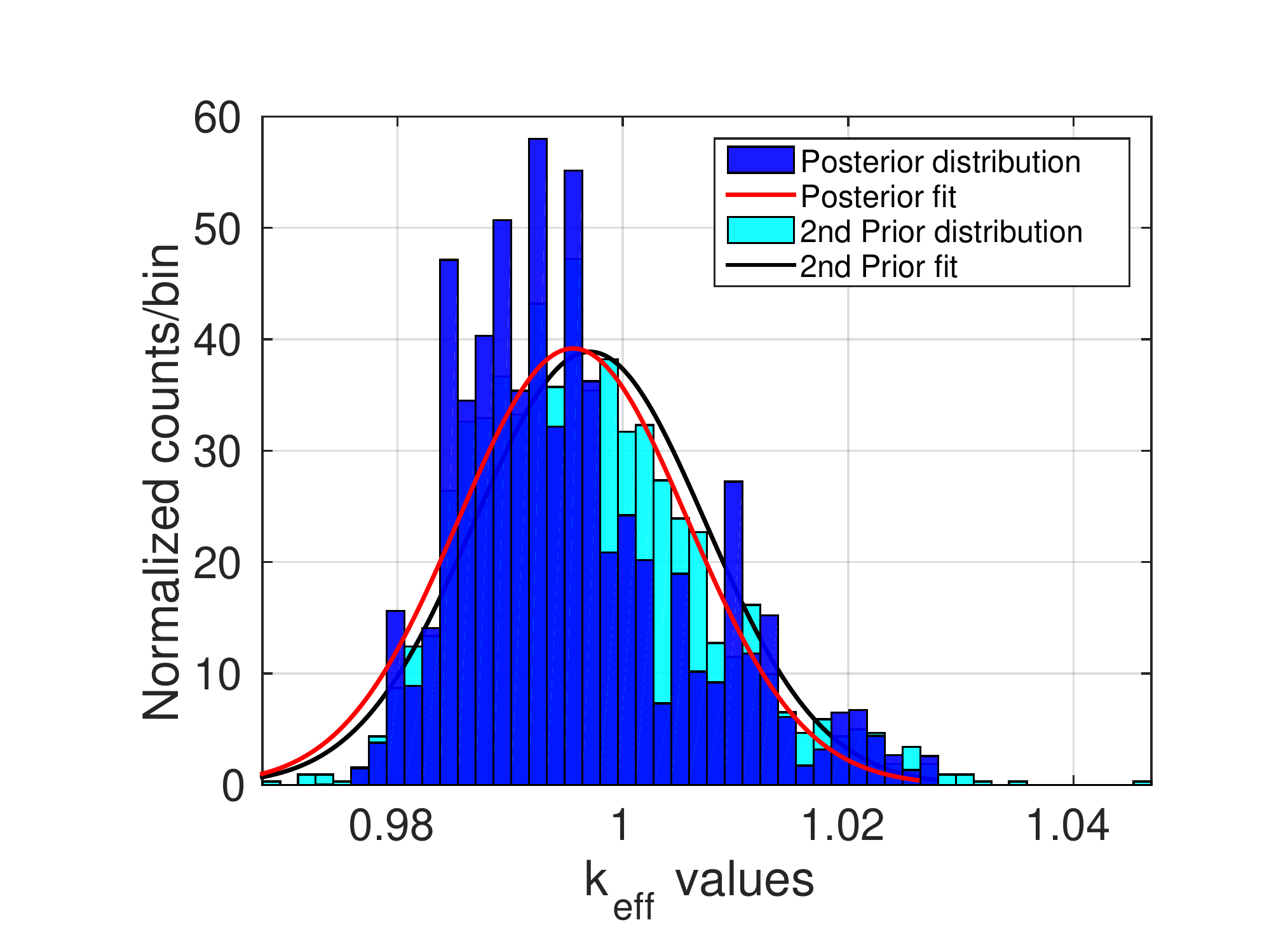} 
   \includegraphics[trim = 5mm 70mm 5mm 60mm, clip, width=0.48\textwidth]{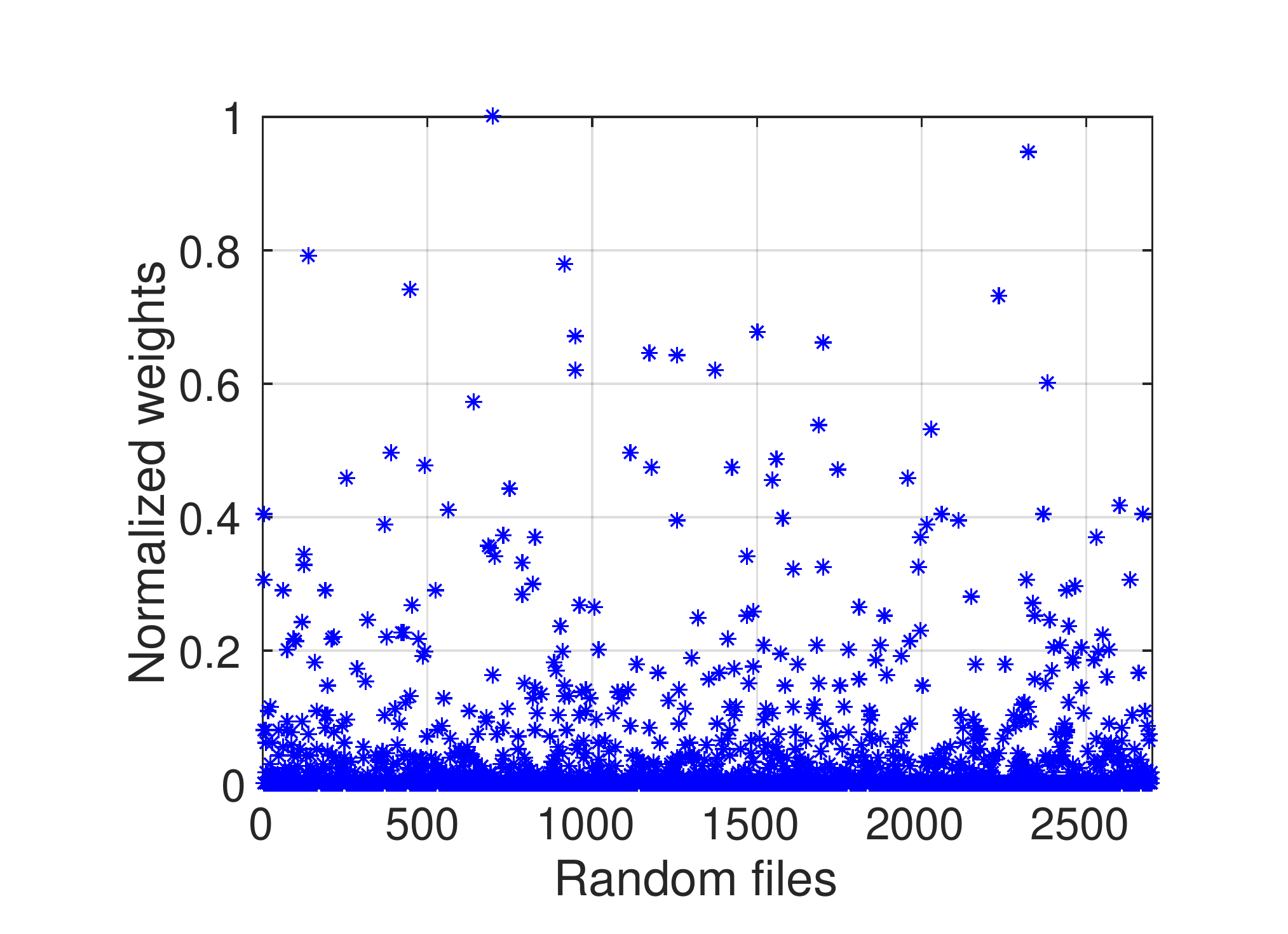} 
   \includegraphics[trim = 5mm 70mm 5mm 60mm, clip, width=0.48\textwidth]{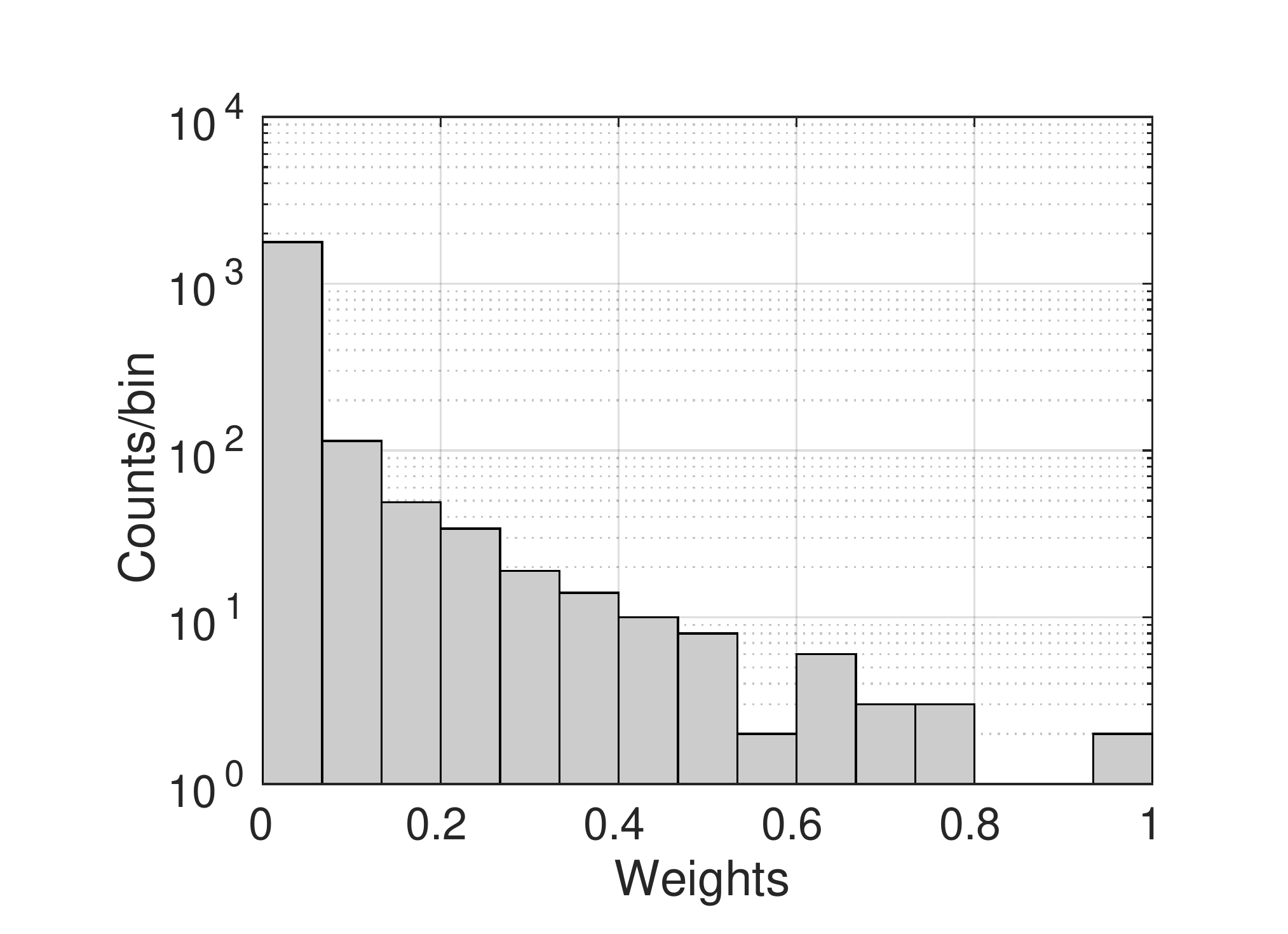} 
  \caption{\textcolor{blue}{The 2nd prior and posterior $k_{\rm eff}$ distributions due to the variation of $^{208}$Pb nuclear data for the hmf57c1 benchmark with the corresponding file weights computed using only selected experiments from EXFOR. Also presented are the convergence plots for the mean $k_{\rm eff}$  and the ND uncertainty of the posterior distribution as well as scatter plot of the weights computed. \textcolor{red}{The 2nd prior represents the $k_{\rm eff}$ distribution after the implementation of the threshold weight presented in Fig.~\ref{fig_BayesUpdate}(note: a total of 2046 random ND files were accepted)}. The error bars on the convergence plot of the ND uncertainty, represent the estimated uncertainty on the ND data uncertainty for each iteration. More details on how to compute this uncertainty has been presented in Refs.~\cite{Alhassan-2014ANE,Helgesson-2013,Sjostrand-2013a}. A Prior ND uncertainty of 1025 pcm and a posterior ND uncertainty of 1018 pcm were obtained. Based on the weights computed, an effective sample size (ESS) of 245 was obtained (i.e. for the case of weighted channels). The weight distribution is given in the log scale.}}
  \label{plot_only_X4}
  \end{figure}

Similarly in Fig.~\ref{plot_only_X4}, the convergence in the mean and the $^{208}$Pb ND uncertainty for the posterior $k_{\rm eff}$ distribution \textcolor{red}{of the 1st update} due to the variation of  $^{208}$Pb nuclear data in the hmf57c1 benchmark is presented. Also presented is a scatter plot of the weights with its corresponding distribution, as well as the corresponding 2nd prior and posterior $k_{\rm eff}$ distributions. \textcolor{red}{The 2nd prior distribution as presented in Fig.~\ref{plot_only_X4} represents the $k_{\rm eff}$ distribution after the implementation of the threshold weight as presented in Fig.~\ref{fig_BayesUpdate}(note: a total of 2046 random ND files were accepted)}. The weights were computed using selected differential experimental data from EXFOR presented ealier in Table~\ref{Exp_data}. From the convergence plot, it can be seen that both the mean and the ND uncertainty converged after a number of iterations - the percentage change in the last five iterations was less than 1\%. It can also be observed that the error bars (which represents the uncertainty on the ND uncertainty estimated) are large but reduces as the number of samples increases with the number of iterations. In all, a total of 2040 accepted random ND files (i.e. accepted files based on the weight threshold criterion presented in Fig.~\ref{fig_BayesUpdate}) were used which resulted in a mean and a ND uncertainty of 0.99707 and 1025$\pm$16 pcm respectively (i.e. in the case where the channels were weighted). The prior ND uncertainty reduced marginally to posterior ND uncertainty of 1018 pcm. This is however different in the case of unweighted channels, where a relatively larger reduction in the posterior ND uncertainty was achieved (ND uncertainty of 986 pcm was obtained). The reduction in ND uncertainty observed, gives an indication that, by including information from differential experimental data in the computation of integral observables such as the $k_{\rm eff}$, the spread in $k_{\rm eff}$ due to  ND uncertainties could be reduced. From Fig.~\ref{plot_only_X4}, it can be observed from the distribution of file weights that, a small number of random ND files were assigned with large weights, above 0.8 for example. This explains why a relatively small reduction in the posterior spread was obtained as seen in the Figure. \textcolor{red}{The few ND files with significant weights obtained are however in agreement with earlier observations made in Refs.~\cite{helgesson-2014incorporating,Helgesson-2017combining}. As noted in Ref.~\cite{Helgesson-2017combining}, if we assume that the models are perfect, a possible explanation could be attributed to our sampling of model parameters from a rather wide parameter space which could lead to a combination of model parameters being drawn from a region of the parameter space where the likelihood is low. A possible solution is then to increase the sample size which could be computationally expensive. Another solution is to  resample model parameters based on the 'best' file but instead, with smaller parameter widths. This could lead to a more even distribution of file weights computed from EXFOR. But since the models are not perfect, another reason could be because of the presence of model defects. The effect of these model defects have not been considered in this work. There are however on-going efforts to include the effect of these defects in modern nuclear data evaluations (See Refs.~\cite{koning-2015bayesianfull,Helgesson-2018treating,Schnabel-2018first,Leeb-2008consistent}).} \textcolor{blue}{Another explanation could be due to the presence of resonance-like spikes in he (n,inl) cross section for example, which were not taken into account in our model calculations since the TALYS code only gives smooth curves for cross sections.} Based on the file weights computed (i.e. in the case of weighted channels), an effective sample size (ESS) of 245 was obtained while an ESS of 127 was obtained for the unweighted channel's case. 

\textcolor{blue}{From the file weights computed for the 1st Bayesian update, a new 'best' file (i.e. file number 0752) was selected. To determine the performance of the selected file, an average $\chi^2$ was computed between the selected file and selected experimental data from the following channels: ((n,tot), (n,non-el), (n,inl), (n,$\gamma$) and the (n,2n)), and compared with average $\chi^2$ values computed for different ND libraries (ENDF/B-VIII.0, JEFF-3.3, JENDL-4.0, CENDL-3.1 and TENDL-2017) using the same experimental data sets and methodology, and presented in Table~\ref{compare_weights11}. The nominal file as presented in the Table represents the prior file before adjustment. \textcolor{blue}{No data was available for the (n,non-el) cross section in the JEFF-3.3 and JENDL-4.0 libraries, therefore, the (n,non-el) cross sections for these libraries were obtained by subtracting the (n,el) from their total cross sections. In the case of the $\chi^2$ results presented for the ENDF/B-VIII.0, JEFF-3.3, JENDL-4.0, TENDL-2017, CENDL-3.1 libraries as well as our nominal file, the channels used were unweighted.}}

\begin{table}[h!]
  \begin{center}
  \centering
  \tabcolsep=0.11cm
  \caption{\label{compare_weights11} \textcolor{blue}{Comparison of different nuclear data libraries as well as adjustments from this work using the reduced chi squared (see Eq.~\ref{gen_chi2}). Note that only experimental data from the (n,tot), (n,non-el), (n,inl), (n,$\gamma$) and the (n,2n) cross sections of $^{208}$Pb were considered. \textcolor{blue}{No data was available for the (n,non-el) channel in the JEFF-3.3 and JENDL-4.0 libraries and therefore, the (n,non-el) was obtained by subtracting the (n,el) from their total cross sections.} As presented in the Table, the 1st update (weighted channels) represents the adjustment with differential data only where each channel was assigned a weight equal to its average cross section, while the (unweighted channels) represents adjustment where all considered channels were assigned with equal weights. \textcolor{red}{In the case of the 2nd update, only the hmf57c1 was used for adjustment while in the case of 'This work (global likelihood)', adjustments were carried out using selected experiments from EXFOR and the hmf57c1 benchmark}. In the case of the global likelihood, both the weighted and unweighted channels gave the same 'best' file. \textcolor{blue}{Note: the $\chi^2$ results given for the ENDF/B-VIII.0, JEFF-3.3, JENDL-4.0, TENDL-2017, CENDL-3.1 and the nominal file were obtained using  unweighted channels.}}}
    \begin{tabular}{lcccccc}
    \toprule
    Libraries & (n,tot) & (n,non-el) &  (n,inl)  & (n,2n) &  (n,$\gamma$)  & Avg $\chi^2$\\
    \midrule 
    ENDF/B-VIII.0  & 4.50 & 0.29 & 19.34  &  2.34  & 1.13  &  5.52  \\
    JEFF-3.3  & 3.16 &   0.16 & 23.03 & 3.82  & 0.37  &  6.11  \\
    JENDL-4.0   &  3.16 &  0.09  &  25.04 &  3.82 & 0.37  & 6.57 \\
    TENDL-2017 & 3.54 &  0.16  &  3.81  &  8.97 &  8.95  &5.08 \\
    CENDL-3.1 & 4.55  & 1.17  &  23.48 & 1.61  &  2.38  & 6.64  \\
    Nominal (prior) file & 4.78 &  0.16  & 17.52  & 2.10   & 8.30  & 6.57 \\
     \pbox{20cm}{This work (1st update) \\ (unweighted channels)} &  5.34  & 0.02  & 10.51 & 4.60  &  3.62 & 4.82  \\
     \pbox{20cm}{This work (1st update) \\ (weighted channels)} & 3.26 & 0.18 & 6.70 & 3.83 & 69.11 & 16.61 \\
   \pbox{20cm}{This work (2nd update) \\ (unweighted channels)} & 8.02  & 6.40 & 5.52 & 39.45  & 70.66 &  26.01 \\
    \pbox{20cm}{This work (global likelihood)}  & 10.57  & 4.22  & 7.94  & 0.88 & 0.83 & 4.89 \\
     \bottomrule
    \end{tabular}
  \end{center}
\end{table}

\textcolor{blue}{From Table~\ref{compare_weights11}, the 1st update (weighted channels) represents the adjustments with only differential data from EXFOR where each channel was assigned a weight equal to its average cross section while the (unweighted channels) represents adjustment where all channels were not weighted. \textcolor{red}{In the case of the 2nd update (as presented in the Table), only the hmf57c1 was used for adjustment while in the case of 'This work (global likelihood)', adjustments were carried out using selected experiments from EXFOR + the hmf57c1 benchmark. More results on the 2nd update and the global (combined) likelihood are presented in sections~\ref{2nd_update} and \ref{global_weights} respectively}. It can be observed from Table~\ref{compare_weights11} that the evaluation from this work (i.e. in the case of 1st update (unweighted) and adjustment with the global likelihood function) out performed the other libraries with respect to the average $\chi^2$. In the case where the 1st update (weighted channels) was compared with its unweighted, it was observed that, the weighted 'best' file out performed the unweighted file for the (n,tot), (n,inl), (n,2n) channels but compared poorly in the case of the average $\chi^2$ and the (n,$\gamma$) channel. This is not surprising since channels with larger weights are given more importance in the random search for the best file than channels with smaller weights. Since the (n,tot), (n,inl), (n,2n) channels have relatively large average cross sections, they were assigned with larger channel weights while the (n,$\gamma$) channel with a relatively smaller average cross section was assigned a smaller channel weight.} \textcolor{red}{One possible reason for the relatively large average $\chi^2$ of 26.01 obtained for the 2nd update (as seen in Table~\ref{compare_weights11}) could be due to over-fitting to integral experimental data since the calculation uncertainties were not taken into account. Also, for a fast system such as the hmf57c1 benchmark, the (n,$\gamma$) channel for example would have minimal effect on the computation of the benchmark $k_{\rm eff}$, hence the large $\chi^2$ value obtained for this channel.} 



\subsection{2nd Bayesian Update: integral benchmarks}
\label{2nd_update}
\begin{figure}[h!] 
  \centering
 \includegraphics[trim = 5mm 72mm 5mm 65mm, clip, width=0.47\textwidth]{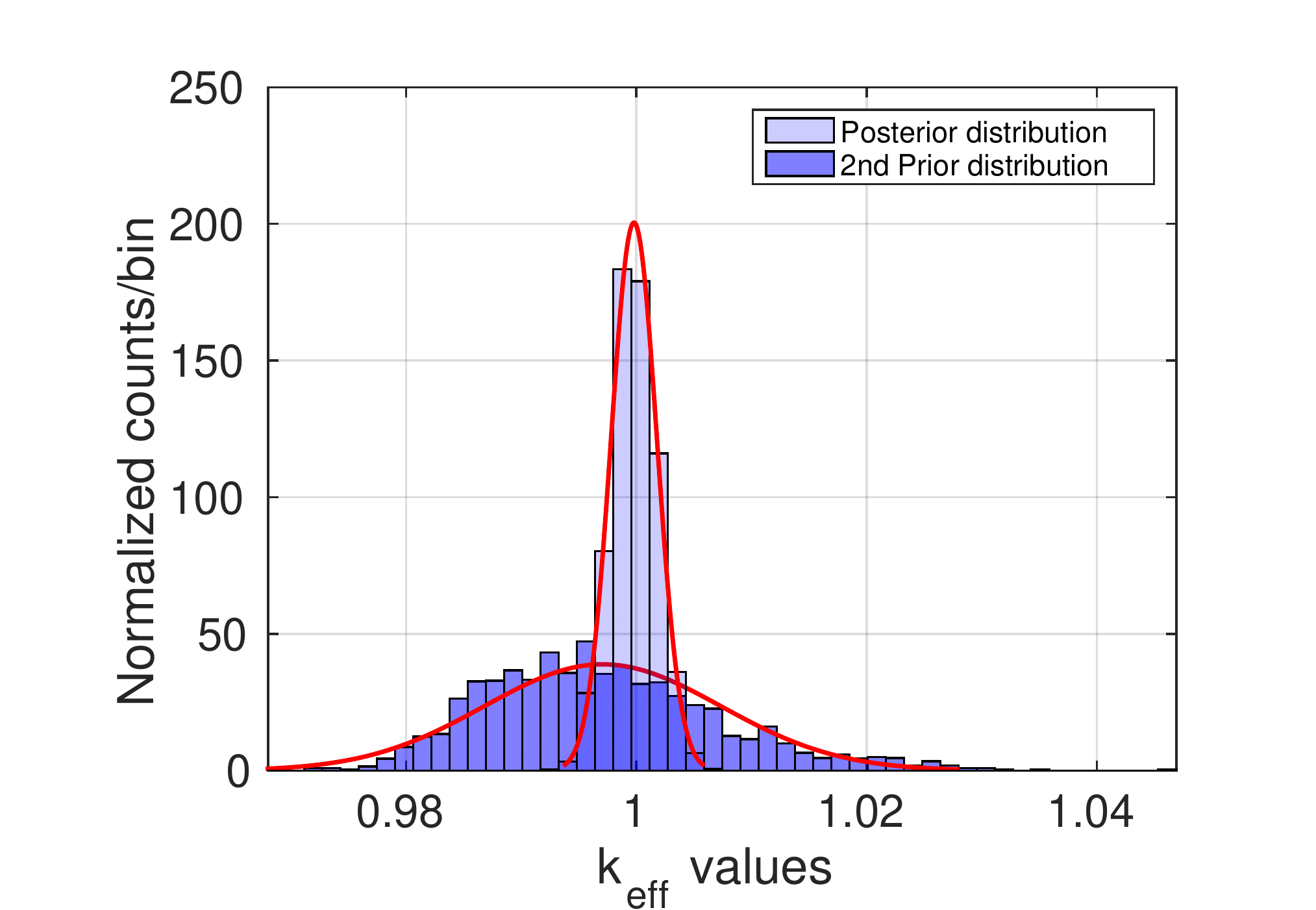}  
 \includegraphics[trim = 5mm 72mm 5mm 65mm, clip, width=0.47\textwidth]{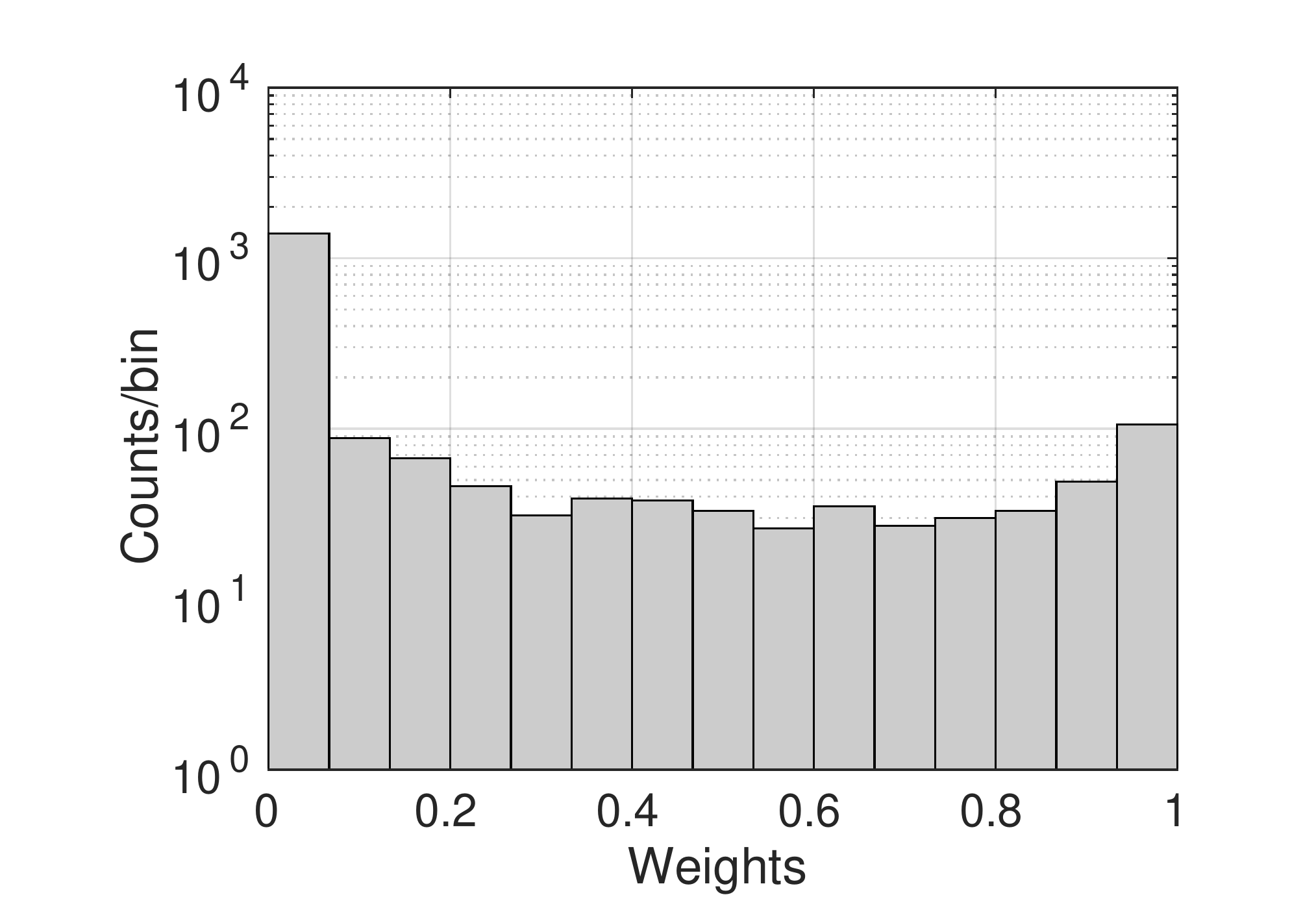} 
   \includegraphics[trim = 5mm 70mm 5mm 60mm, clip, width=0.47\textwidth]{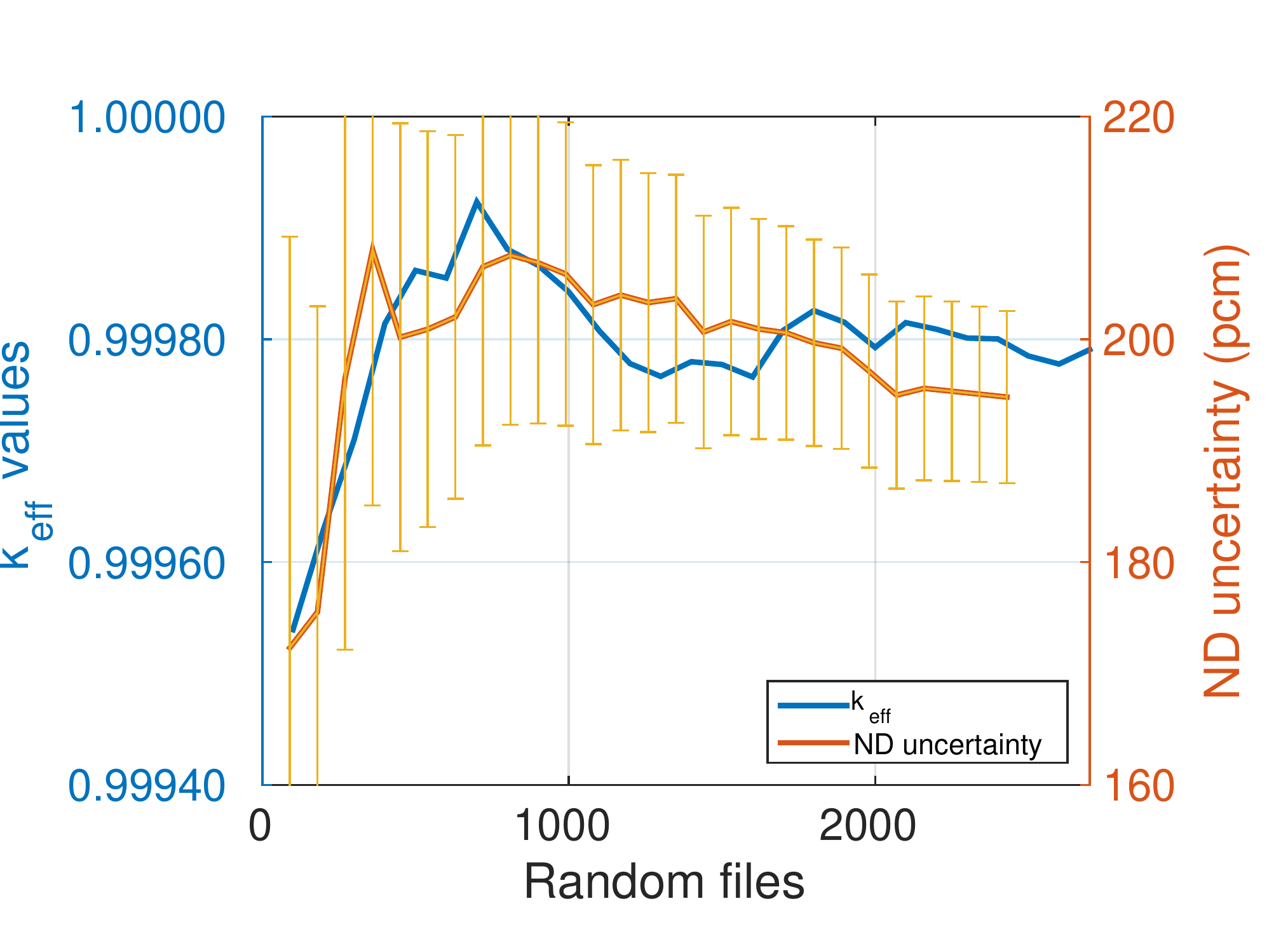} 
     \includegraphics[trim = 5mm 70mm 5mm 60mm, clip, width=0.47\textwidth]{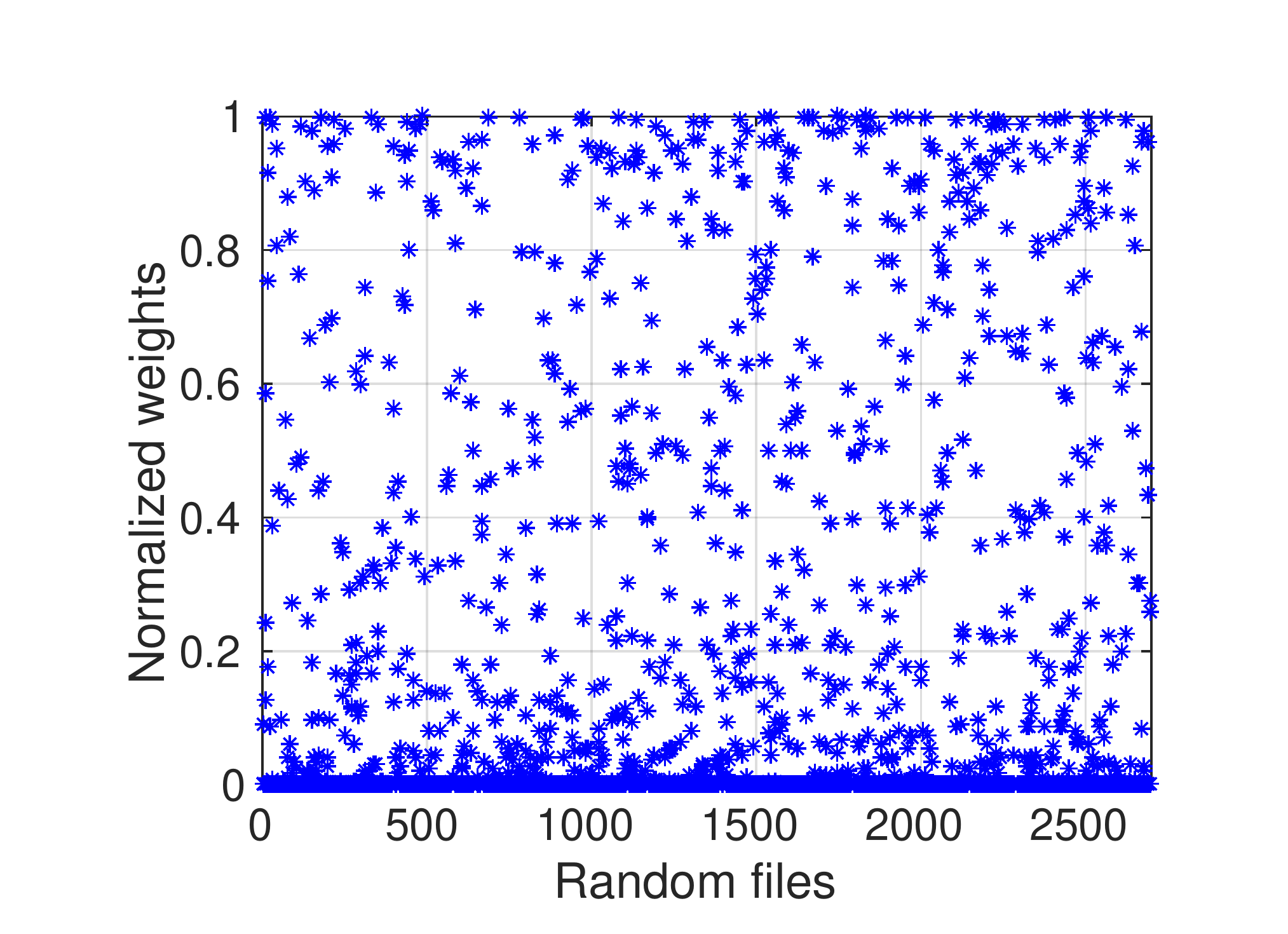} 
  \caption{Prior (2nd) and posterior $k_{\rm eff}$ distributions due to the variation of $^{208}$Pb nuclear data for the hmf57c1 benchmark with its corresponding weight distribution (and scatter plot) computed using hmf57c1 benchmark experimental information. Also presented is the convergence in the mean and ND uncertainty of  the posterior distribution. The error bars on the convergence plot of the ND uncertainty represent the estimated uncertainty on the ND data uncertainty for each iteration. The prior and posterior $k_{\rm eff}$ distributions are fitted with Gaussian distributions for the purpose of eye guidance only. An average statistical uncertainty of 40 pcm was recorded for the benchmark. The weight distribution is given in the log scale.}
  \label{fig_keff_Pb208_hmf57c1}
 \end{figure}

\textcolor{red}{The 2nd prior (as presented in the 1st update) was used as the prior distribution for the 2nd update. The prior distribution here then represents $k_{\rm eff}$ distributions using the ND files that were accepted (or passed the weight threshold test presented earlier in Fig.~\ref{fig_BayesUpdate}).} \textcolor{blue}{In Fig.~\ref{fig_keff_Pb208_hmf57c1}, the prior (2nd) and posterior $k_{\rm eff}$ distributions due to the variation of $^{208}$Pb nuclear data for the hmf57c1 benchmark with its corresponding weight distribution computed using only experimental information from the hmf57c1 benchmark is presented. Also presented is the convergence plot of the mean and ND uncertainty of the posterior $k_{\rm eff}$ distribution as well as a scatter plot of the weights.}
\textcolor{blue}{Similar to Fig.~\ref{plot_only_X4}, the fluctuations observed in the mean and the ND uncertainty were observed to be small (less than 1\% (relative difference) was obtained between the last 5 iterations). Large error bars can be observed for the convergence of the ND uncertainty especially in the case of smaller sample sizes which reduces as the number of iterations (and sample size) increases. From Fig.~\ref{fig_keff_Pb208_hmf57c1}, it can be observed that, the large prior uncertainty of 1062 pcm was reduced to a posterior uncertainty of 194 pcm representing an uncertainty reduction of 82\%. Also, the mean of the posterior distribution of 0.99981 obtained is close to the experimental $k_{\rm eff}$ as desired. It should however be noted that, in this particular case, only the experimental benchmark uncertainty information was used in the computation of the file weights, hence the very narrow spread observed in the posterior distributions. As noted earlier, in Ref.~\cite{Alhassan-2015NDreduction} an additional uncertainty term (uncertainty in the calculation due to nuclear data) was included in the computation of file weights which resulted in much larger combined benchmark uncertainties and therefore larger posterior distributions. This approach was however not used in this work.}

Additionally, it should be noted that, the $\chi^2$  values (and therefore the file weights) in the 2nd update were computed with only experimental data from the hmf57c1 benchmark and therefore, even though an updated $k_{\rm eff}$ = 1.00002 (see Table~\ref{Exp_data_Newresults}) was obtained for the new 'best' file (i.e. for the 2nd update), the file performed poorly when compared back with differential experimental data from EXFOR (see $\chi^2$ values presented in Table~\ref{compare_weights11}). An average $\chi^2$ value of 26.01 was obtained for the new 'best' file (2nd update), compared with 4.82 from the 1st update (unweighted channels), and 5.52, 6.11, 6.57, 5.08 and 6.64 for the ENDF/B-VIII.0, JEFF-3.3, JENDL-4.0, TENDL-2017 and  CENDL-3.1 libraries respectively. The combined adjustment using the likelihoods from the two updates (i.e. the global likelihood (unweighted)), however, gave an average $\chi^2$  of 4.89. This gives an indication that by combining information from both differential and integral experiments, improvements on the nominal (prior) file can be achieved. 

\textcolor{blue}{Table~\ref{prior_post_NDuncert} presents a summary of results of the mean $k_{\rm eff}$ with corresponding ND uncertainties, and the Effective Sampling Sizes (ESS) for the 1st and 2nd prior distributions as well as for the posterior distributions of the 1st, 2nd and combined updates. The 1st prior represents the $k_{\rm eff}$ distribution of random nuclear data files before the implementation of the threshold weight (i.e. with a total of 2700 random ND files) while the 2nd prior represents the distribution after the implementation of the threshold weight (a total of 2046 random ND files were accepted) as presented in Fig.~\ref{fig_BayesUpdate}. From the Table, it can be observed that the ESS for the (unweighted case) are relatively smaller than the weighted case for the posterior distributions. This could be attributed to the large weights assigned to the channels with large average cross sections such as the (n,tot), (n,inl) and the (n,non-el) channels and which also, TALYS-1.6 was able to reproduce within experimental uncertainties. It was however observed that, TALYS-1.6 was not able to reproduce the (n,$\gamma$) cross sections of $^{208}$Pb (which also has a relatively smaller average cross section) for a large number of the random cross sections produced.}

In the case of the posterior (combined) as presented in the Table, small ESS of 20 and 73 were obtained for the unweighted and weighted channels respectively. The small ESS obtained, could be attributed to the rather few files assigned with large weights which resulted in larger differences in file weights for some random ND files. Also, smaller ESS results in unconvergence of the distribution under consideration since it implies that very few files have significant impact on the posterior distribution. In order to achieve a more even weight distribution and hence higher ESS values, outlier ND files could have been discarded and the file weights renormalized with the maximum weight. This however, must be done with caution as extremely good 'outlier' ND files with large weights may be discarded. Since the main objective of the combined adjustment is to identify the ND file that reproduces both differential and integral data, the outlier ND files were not discarded. 

\begin{table}[h!]
  \begin{center}
  \footnotesize
  \centering
  \tabcolsep=0.11cm
    \caption{\label{prior_post_NDuncert} \textcolor{blue}{Summary of results for the mean $k_{\rm eff}$ with corresponding nuclear data uncertainty ($\pm$ uncertainty on the estimated ND uncertainty) and the Effective Sampling Size (ESS) for the 1st and 2nd priors, and the posterior distributions of the 1st, 2nd and the combined updates. Weighted channels represents the case where each channel was assigned a weight equal to its average cross section over the considered energy range while the unweighted channels represents the case where all channels were assigned equal weights. The 1st prior here represents the distribution of the $k_{\rm eff}$ without the inclusion of experimental data while in the case of the 2nd prior, the information from the differential experimental data were used to exclude some random files based on a weight threshold (see Fig.~\ref{fig_BayesUpdate}). }}
    \begin{tabular}{lccc|ccc}
    \hline
  & \multicolumn{2}{c}{Unweighted channels} & & \multicolumn{2}{c}{Weighted channels} & \\
    \hline
     Distributions & Mean $k_{\rm eff}$ &  \pbox{20cm}{ND uncertainty \\ (pcm)} & ESS & $k_{\rm eff}$ & \pbox{20cm}{ND uncertainty \\ (pcm)}  & ESS  \\
    \hline
  1st prior & 0.99841 & 1103$\pm$15 &  2700 & 0.99841 & 1103$\pm$15 &  2700  \\
  2nd Prior &  0.99749 & 1065$\pm$17 & 2048 & 0.99707 &  1025$\pm$16 & 2038 \\
  Posterior (1st update) &  0.99546 & 986$\pm$119 & 127 & 0.99558 & 1018$\pm$87 & 245 \\
  Posterior (2nd update) & 0.99981 & 194$\pm$8 & 475 & 0.99979 & 195$\pm$8 & 492\\
  Posterior (Combined) & 0.99988 & 164$\pm$56 & 20 & 0.99941 & 203$\pm$55  & 73\\
    \hline
    \end{tabular}
  \end{center}
\end{table}

\subsection{Using the global likelihood function: EXFOR + hmf57c1 benchmark}
\label{global_weights}
\begin{figure}[h!] 
  \centering
   \includegraphics[trim = 15mm 12mm 15mm 18mm, clip, width=0.49\textwidth]{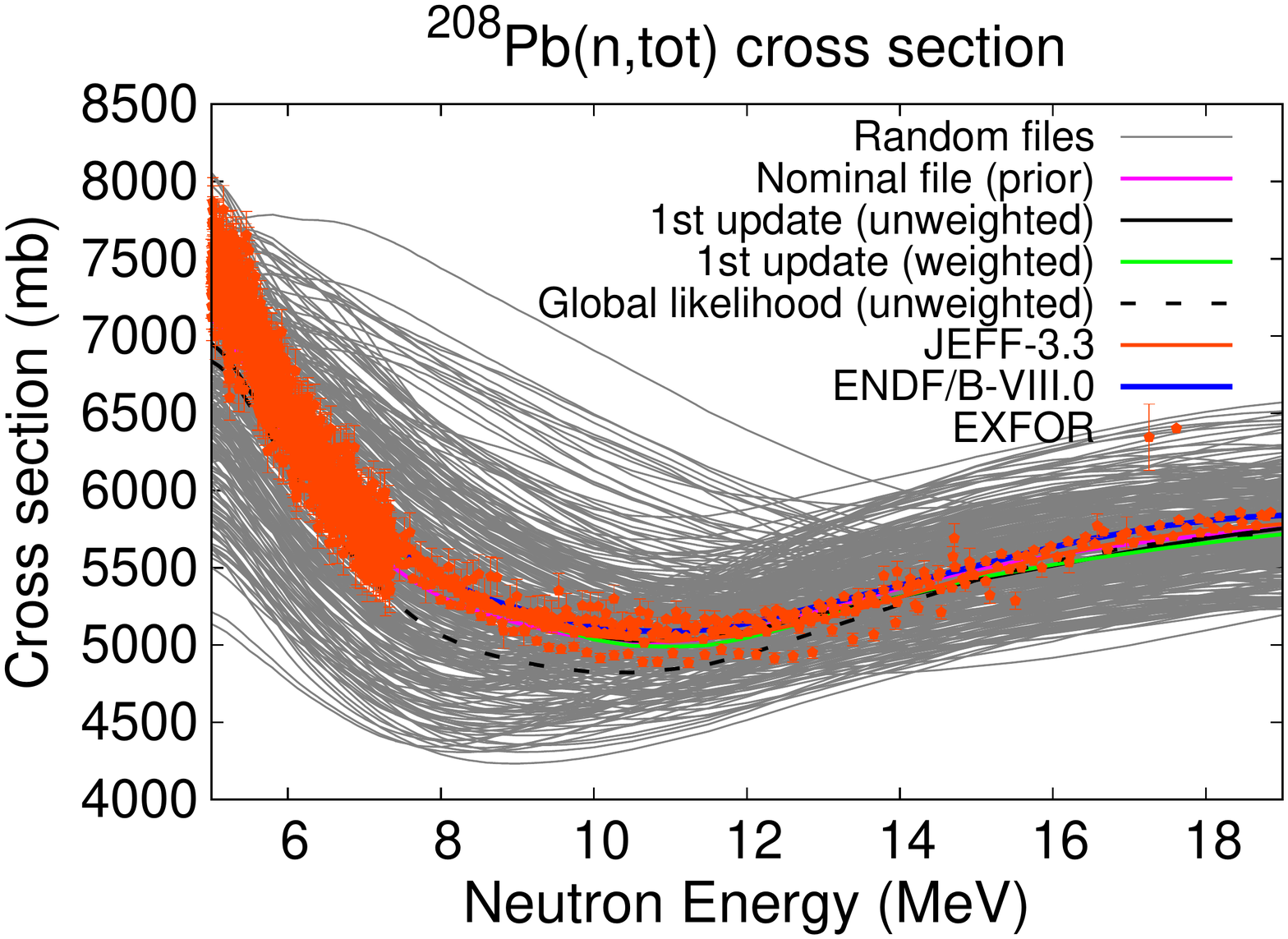} 
   \includegraphics[trim = 15mm 12mm 15mm 18mm, clip, width=0.49\textwidth]{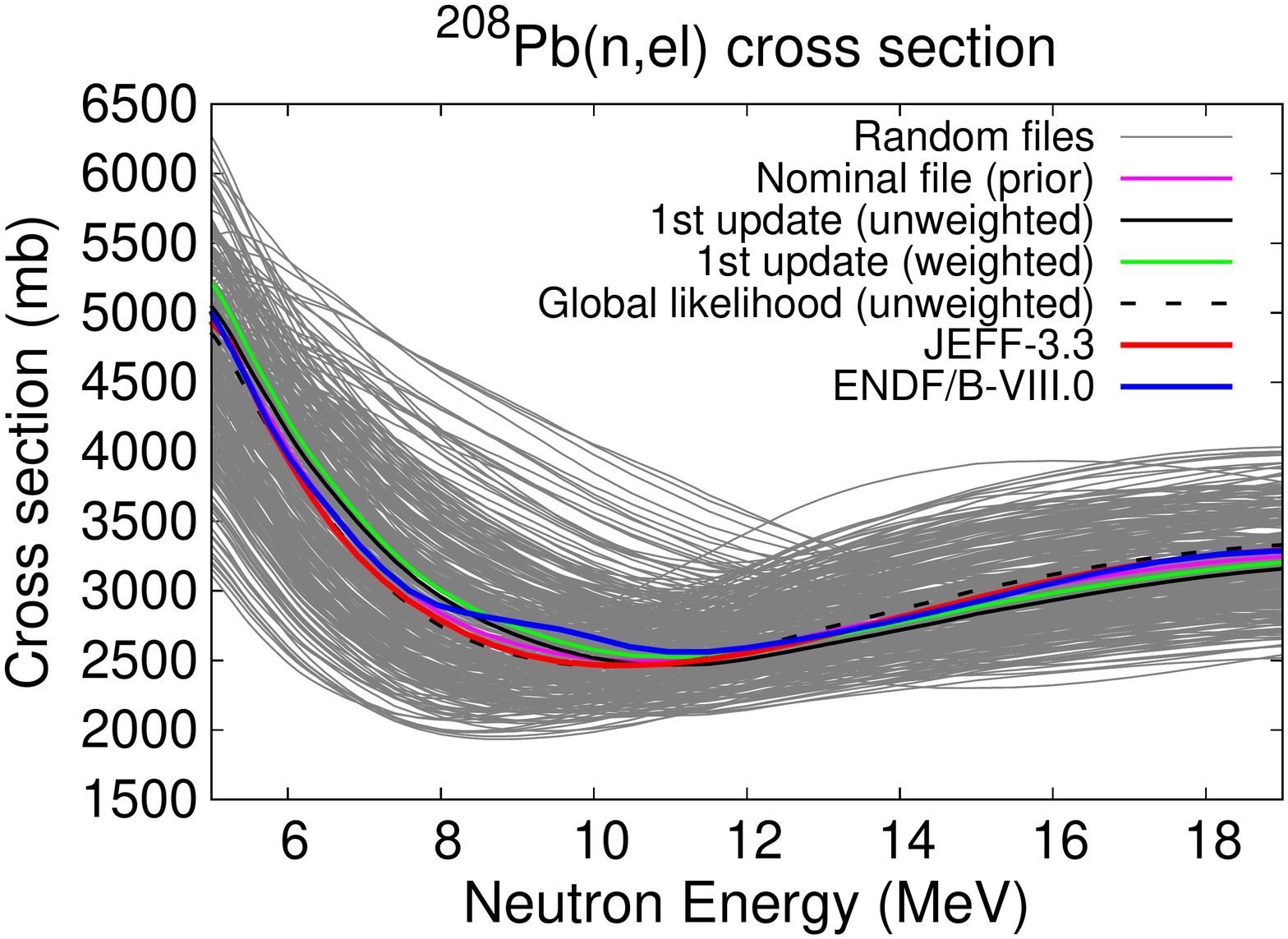} 
   \includegraphics[trim = 15mm 12mm 15mm 18mm, clip, width=0.49\textwidth]{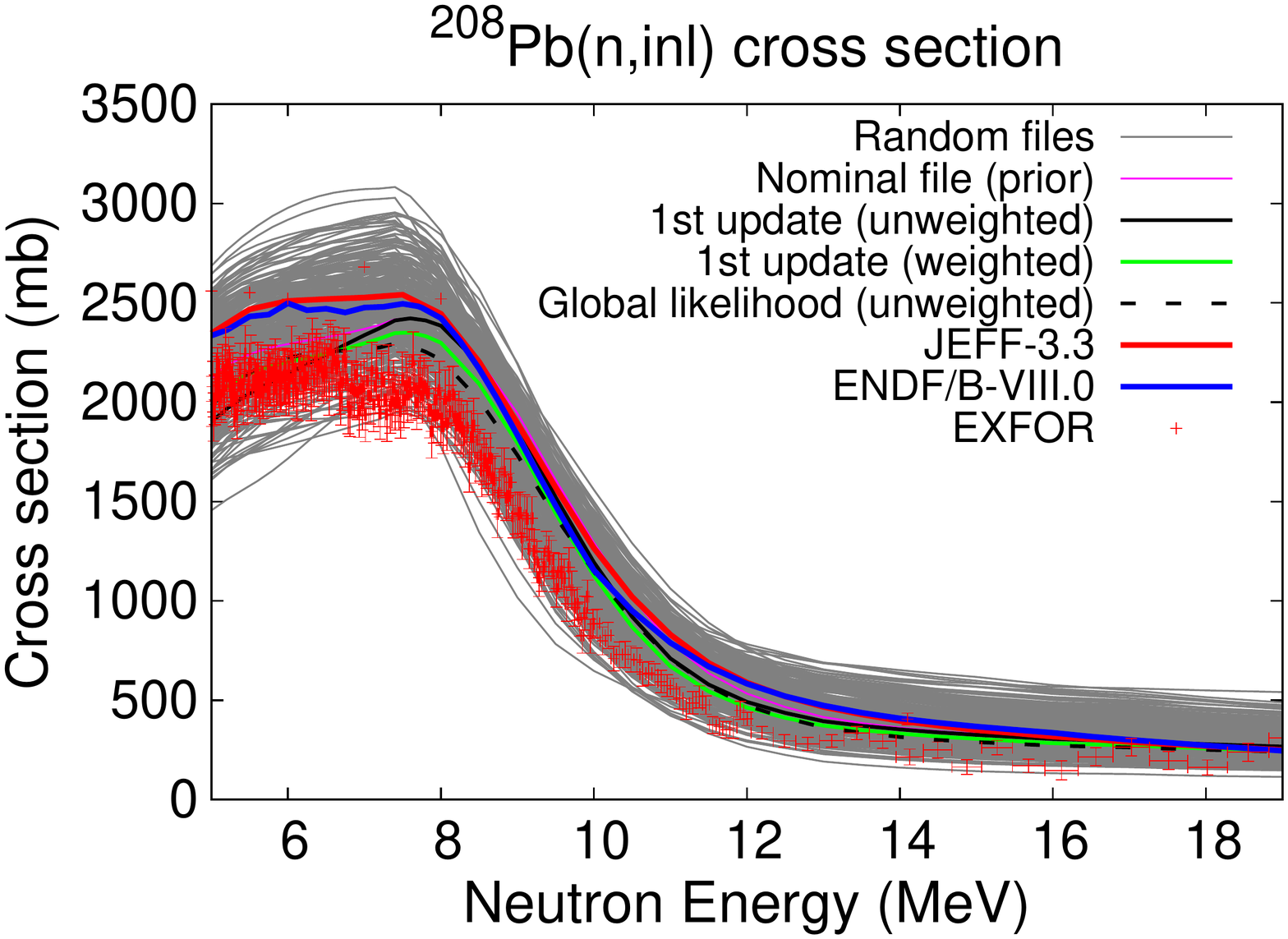} 
   \includegraphics[trim = 15mm 12mm 15mm 18mm, clip, width=0.49\textwidth]{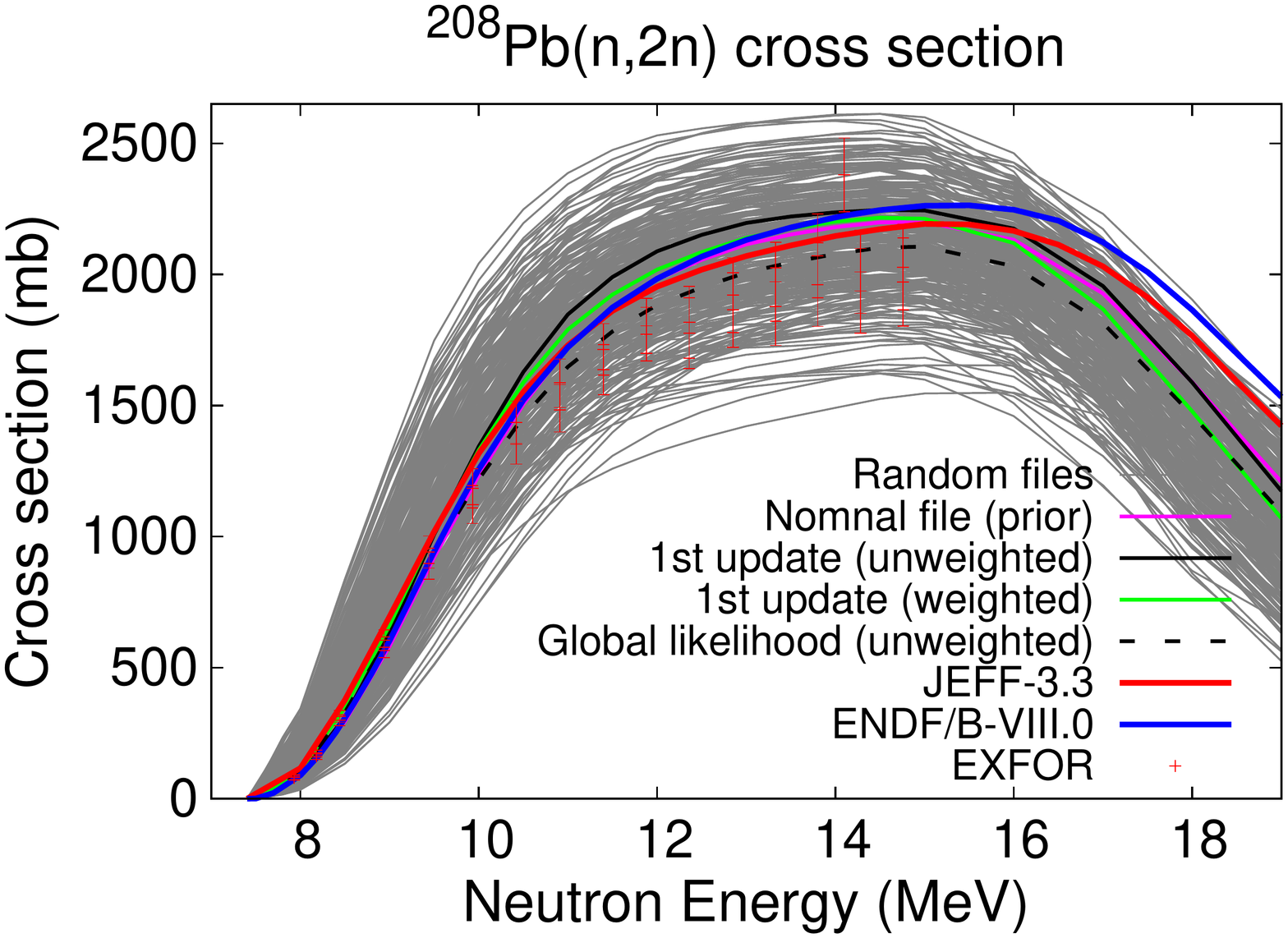} 
  \caption{Comparison of file performance between this work and the evaluations from the ENDF/B-VIII.0 and JEFF-3.3 ND libraries and differential experimental data from EXFOR (between 5 to 20 MEV) for $^{208}$Pb (n,2n), (n,el), (n,inl) and (n,2n) cross sections. \textcolor{blue}{The nominal file (prior) is the ND file around which the other random files were generated. The authors of the different experiments from EXFOR have been lumped together and labelled as EXFOR. The weighted and unweighted represents cases where channels were weighted with their average cross sections or where all channels carried equal weights respectively.}}
  \label{coma_exfor_bench}
  \end{figure}

\begin{figure}[h!] 
  \centering
   \includegraphics[trim = 15mm 12mm 15mm 18mm, clip, width=0.49\textwidth]{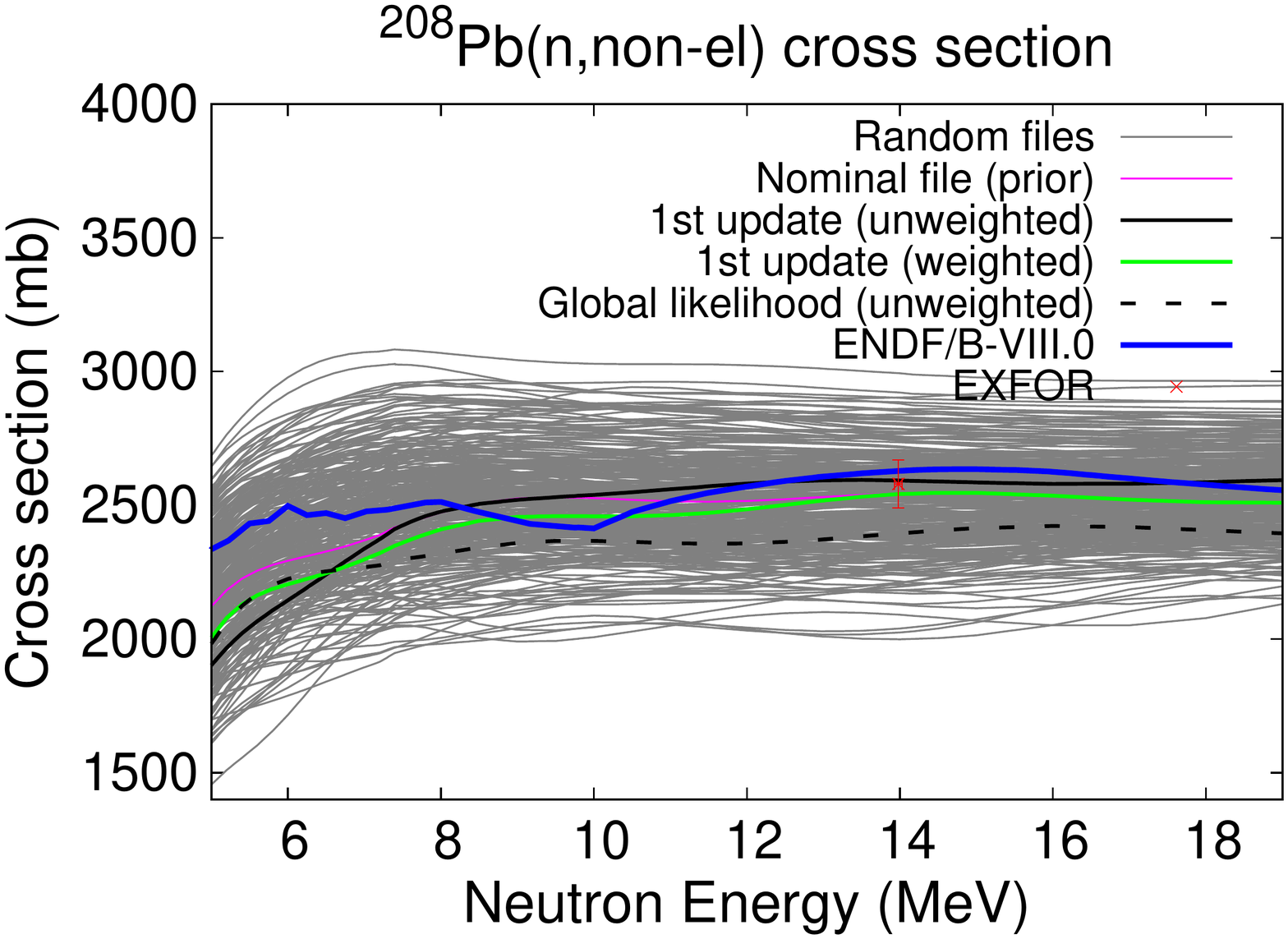} 
   \includegraphics[trim = 15mm 12mm 15mm 18mm, clip, width=0.49\textwidth]{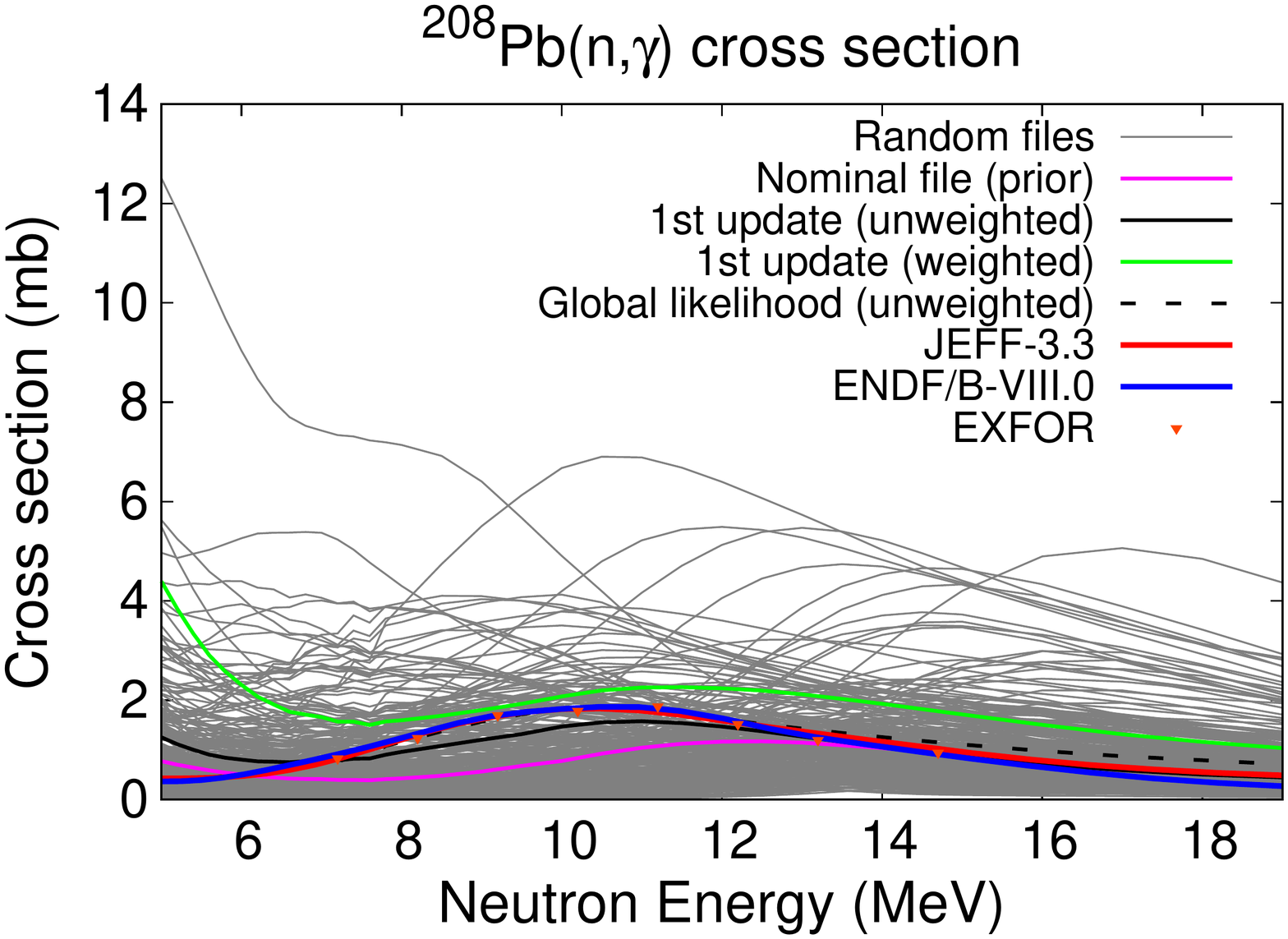} 
  \caption{Comparison of file performance between this work and the major nuclear data libraries as well as with differential experimental data from EXFOR (between 5 to 20 MEV) for $^{208}$Pb (n,non) and (n,$\gamma$) cross sections. \textcolor{blue}{The nominal file (prior) is the file around which the other random files were generated. The authors of the different experiments from EXFOR are not presented, instead, they have all been labelled as EXFOR. The weighted and unweighted represents cases where channels were weighted with their average cross sections or where all channels carried equal weights respectively.}}
  \label{coma_exfor_bench2}
  \end{figure}

\textcolor{blue}{A final test of an adjustment is to compare the final adjusted file back with differential experimental data from EXFOR as well as with relevant benchmarks in order to determine its performance. Additionally, comparisons are made with other existing evaluations. In Figs.~\ref{coma_exfor_bench} and \ref{coma_exfor_bench2}, the performance of the adjustments from this work are compared with available differential experimental data (between 5 to 20 MeV) from EXFOR for the (n,tot), (n,el), (n,inl), (n,2n), (non-el) and (n,$\gamma$) cross sections of $^{208}$Pb as well as with evaluations from the ENDF/B-VIII.0 and JEFF-3.3 nuclear data libraries. Since no experimental data were available in EXFOR for the (n,el) cross section, we have only compared our evaluations with the ENDF/B-VIII.0 and JEFF-3.3 nuclear data libraries in the case of the (n,el) cross section. The nominal (prior) file is the prior file around which the other random ND files were generated. As seen from the Figs.~\ref{coma_exfor_bench} and \ref{coma_exfor_bench2}, the individual authors of the different experimental sets from EXFOR have not been listed, instead, they have been been lumped together and named as EXFOR. The weighted and unweighted as presented in the Figures, represents cases where channels were weighted with their average cross sections or where all channels carried equal weights respectively.} 

\textcolor{blue}{It can be seen from Fig.~\ref{coma_exfor_bench} (in the case of the (n,tot) cross section) that, the 'best' file from the combined adjustment using the global likelihood function (EXFOR + hmf57c1) deviates slightly from experimental data between about 7 to 12 MeV; this explains the relatively large $\chi^2$ of 10.57 obtained for the (n,tot) cross section in Table~\ref{compare_weights11}. A deviation from experimental data is also observed in the (n,non-el) channel as can be seen in Fig.~\ref{coma_exfor_bench2}. Since the (n,tot) is a sum of the (n,el) with the (n,non-el) cross sections, the deviation observed in the (n,non-el) channel could be a contributory factor to the deviation observed in the (n,tot) channel. It should however be noted that, this interpretation must be made with caution since only one experimental measurement was available for the (n,non-el) channel in EXFOR at the time when this work was carried out. Our 'best' file however outperforms the other evaluations for the (n,2n) and the (n,inl) cross sections as can be seen from the Fig.~\ref{coma_exfor_bench}. The evaluations from JEFF-3.3 and JENDL-4.0 overlap each other for the (n,2n), (n,$\gamma$) and (n,tot) cross sections and therefore the evaluation from JENDL-4.0 has not be shown.}

\textcolor{blue}{In order to determine how our adjusted files compared with integral experiments, the adjusted files (from this work) were inserted into the ENDF/B-VII.0 library and used for criticality calculations for selected benchmarks. The results obtained were compared with benchmark experimental data (see Table~\ref{table_critB}) as well as with calculational results obtained using only the ENDF/B-VII.0 ND library and presented in Table~\ref{Exp_data_Newresults}. ENDF/B-VII.0 was used because it was the library version that came with the version of the MCNPX code used for criticality calculations in this work. In Table~\ref{Exp_data_Newresults}, the ratios of calculated to experimental values (C/E) for the selected lead sensitive benchmarks are also presented. In the case of criticality calculations for each benchmark as presented in the Table, the the ENDF/B-VII.0 library was maintained as the reference (or base) library for all isotopes while only the nuclear data of  $^{208}$Pb was varied. In the case of ENDF/B-VII.0  (column 6 of Table) however, the nuclear data for all isotopes were maintained as the ENDF/B-VII.0 nuclear data library. As mentioned earlier, only the hmf57c1 benchmark was used for adjustment in the 2nd update; all other benchmarks as presented in the Table were used for testing and validation purposes. This was done in order that the same benchmark used for adjustment was not also used for testing and validation. Consequently, three different 'best' files were obtained: (1) adjustment with differential data only (1st update), (2) adjustment with the hmf57c1 benchmark only (2nd update), and (3) the combined adjustments using the combined (global) likelihood function.} 

\begin{table}[h!]
  \begin{center}
  \footnotesize
  \centering
  \tabcolsep=0.11cm
    \caption{\label{Exp_data_Newresults} The ratios of calculated to experimental values (C/E) for selected benchmarks for adjustments with differential experimental data only (1st update), the hmf57c1 benchmark (2nd update), and the combined (global) likelihood function (EXFOR +  hmf57c1 benchmark). For the adjustments with hmf57c1, the ENDF/B-VII.0 was used as the reference (base) library for all isotopes, except for $^{208}$Pb which was varied. In the case of ENDF/B-VII.0 (column 6), all isotopes were maintained as the ENDF/B-VII.0 nuclear data library. The benchmark experimental $k_{\rm eff}$ values have been presented previously in Table~\ref{table_critB}.}
    \begin{tabular}{lcccccc}
    \toprule
     Benchmarks & \pbox{20cm}{1st update \\ (unweighted \\ channels)}  & \pbox{20cm}{1st update \\ (weighted \\ channels)}  & \pbox{20cm}{2nd update \\ (hmf57c1)} & \pbox{20cm}{Global likelihood \\ (EXFOR + hmf57c1)} & ENDF/B-VII.0  \\
    \midrule
   \emph{ hmf57c1} &  1.01285 &  0.99629  & 1.00002 &  1.00042 & 0.98959 \\
    hmf57c2 & 1.01633  & 1.00439  & 1.00851   & 1.00201 & 0.99888   &    \\ 
    hmf57c3 & 1.04167 & 1.02459 & 1.03023   & 1.02918 & 1.01726    &  \\ 
    hmf57c4 & 1.00462 & 0.99293  & 0.99793   & 0.99565  & 0.98784    &  \\
    hmf57c5   & 1.04884 & 1.02934 & 1.03580   & 1.03437 & 1.02188   & \\ 
    hmf57c6 & 1.02047 & 1.00322  & 1.00909   &  1.00769 & 0.99713   & \\
    
    hmf27c1  & 1.01150 & 1.00363  & 1.00676   & 1.00558 &  1.00182   &  \\
    pmf35c1  & 1.00891 &  1.00015 & 1.00635   & 1.00290 &  0.99856   &  \\
    hmf64c1  & 1.02110  & 1.00237 & 1.00936  & 1.00263 & 0.99455   & \\
    hmf64c2  & 1.02852& 1.00532 & 1.01265   & 1.01058  & 0.99613  & \\
    \bottomrule
    \end{tabular}
  \end{center}
\end{table}
\textcolor{blue}{From Table~\ref{Exp_data_Newresults}, it can be observed that the adjustment from the 1st update (weighted channels), performed relatively well when compared with its unweighted for most of the benchmarks. This could be because (as observed earlier in Table~\ref{compare_weights11}), relatively smaller $\chi^2$ values were obtained (i.e. in the case of the 1st update (weighted channels)) for the (n,tot), (n,inl) and (n,2n) cross sections, which are relatively important cross sections for fast systems. Fast systems such as the hmf57c1 benchmark have small capture reactions and hence, though the (n,$\gamma$) cross section ((in the case of weighted channels) performed badly with respect to the $\chi^2$ presented in Table~\ref{compare_weights11}, its impact on the calculated $k_{\rm eff}$ was minimal. This explains why even though the best file from the 2nd update (i.e. random file number 1835 with a $k_{\rm eff}$ = 1.00002 for the hmf57c1 benchmark) performed relatively better than the adjustment from the 1st update for most of the benchmarks, the file performed poorly when compared with differential experimental data (see Table~\ref{compare_weights11}). From the combined adjustments using the global likelihood function (EXFOR + hmf57c1), it can be observed that, the adjusted file performs relatively well against all the benchmarks as well as against the ENDF/B-VII.0 library. Furthermore, as seen previously from Table~\ref{compare_weights11}, the best file from the combined adjustment performed quite well against the other nuclear data libraries for the selected cross sections presented.}

As mentioned earlier, in the 2nd update only one benchmark was used for the adjustment, however, a global adjustment with a single benchmark (as carried out in this work), could lead to a global fit to the experimental $k_{\rm eff}$ for the particular benchmark used for adjustment but could have discrepancies when compared with other benchmarks or with experimental data from EXFOR. Therefore, the use of multiple benchmarks (including their correlations) for adjustments has been proposed and is presented in a dedicated paper~\cite{Alhassan-2018Multiple}. Further, a natural extension of this work is to utilized the global (combined) likelihood function computed for the reduction of ND uncertainty in applications as presented in Ref.~\cite{Alhassan-2015NDreduction}. Also, the inclusion of model defects in the adjustment procedure is proposed for future work.

\section{Conclusion}
\label{sect::conclusion}
\textcolor{blue}{A method was presented for combining differential and integral benchmark experimental data for nuclear data adjustments and multi-level uncertainty propagation within the TMC method. The method combines individual likelihood functions computed from two Bayesian updates into a global (combined) likelihood function. The proposed method was applied for the adjustment of neutron induced reactions on $^{208}$Pb in the fast energy region below 20 MeV. The results from the adjustments were compared with available experimental data from EXFOR, the major nuclear data libraries and against a selected number of criticality benchmarks and found to compare quite favourably. In order to take full advantage of relevant integral experiments available, the use of multiple benchmarks for adjustments (including their correlations) and the inclusion of benchmark calculation uncertainties as well as model defects, is proposed for future work.}

\section*{ACKNOWLEDGMENTS}
We would like to thank Petter Helgesson, Stephan Pomp and Georg Schnabel from Uppsala University, Sweden, and David R. Novog from McMaster University, Canada for insightful comments and discussions on this topic. 

\bibliographystyle{physor2014}
\bibliography{references}

\end{document}